\documentclass[aps,prd,reprint,nofootinbib,longbibliography,preprintnumbers]{revtex4-1}

\usepackage{amsmath}
\usepackage{mathtools}
\usepackage{graphicx}
\usepackage[normalem]{ulem}
\usepackage[com pat=1.1.0]{tikz-feynman}
\usepackage{tikz}
\usepackage{aas_macros}
\usetikzlibrary{external}
\usepackage{comment}
\usetikzlibrary{arrows}
\usepackage[colorlinks=true, allcolors=blue]{hyperref}

\newcommand\barparena[1]{\overset{%
   \scriptscriptstyle(-)}{#1}}

\begin{document}

\preprint{LA-UR-23-34188}

\title{BGK subgrid model for neutrino quantum kinetics}

\author{Hiroki Nagakura}
\email{hiroki.nagakura@nao.ac.jp}
\affiliation{Division of Science, National Astronomical Observatory of Japan, 2-21-1 Osawa, Mitaka, Tokyo 181-8588, Japan}
\author{Lucas Johns}
\affiliation{Theoretical Division, Los Alamos National Laboratory, Los Alamos, NM 87545, USA}
\author{Masamichi Zaizen}
\affiliation{Faculty of Science and Engineering, Waseda University, Tokyo 169-8555, Japan}

\begin{abstract}
We present a new subgrid model for neutrino quantum kinetics, which is primarily designed to incorporate effects of collective neutrino oscillations into neutrino-radiation-hydrodynamic simulations for core-collapse supernovae and mergers of compact objects. We approximate the neutrino oscillation term in quantum kinetic equation by Bhatnagar–Gross–Krook (BGK) relaxation-time prescription, and the transport equation is directly applicable for classical neutrino transport schemes. The BGK model is motivated by recent theoretical indications that non-linear phases of collective neutrino oscillations settle into quasi-steady structures. We explicitly provide basic equations of the BGK subgrid model for both multi-angle and moment-based neutrino transport to facilitate the implementation of the subgrid model in the existing neutrino transport schemes. We also show the capability of our BGK subgrid model by comparing to fully quantum kinetic simulations for fast neutrino-flavor conversion. We find that the overall properties can be well reproduced in the subgrid model; the error of angular-averaged survival probability of neutrinos is within $\sim 20 \%$. By identifying the source of error, we also discuss perspectives to improve the accuracy of the subgrid model.
\end{abstract}
\maketitle

\section{Introduction}\label{sec:intro}
Astrophysical phenomena usually involve intricately intertwined multiphysics. Direct numerical simulation is an effective tool to study the physical mechanism behind these complex phenomena, and also to provide theoretical models for interpretations of observed data. Ofttimes, however, the temporal- and spatial scales among different physical processes span many orders of magnitudes, rendering the first-principles simulations prohibitively computationally expensive. This exhibits the need for approximations or coarse-grained approaches.

It has been recognized for many years that neutrino quantum kinetics in core-collapse supernova (CCSN) and mergers of compact objects represented by binary neutron star merger (BNSM) corresponds to such a problem requiring coarse-grained treatments (see reviews in \cite{2010ARNPS..60..569D,2021ARNPS..71..165T,2022arXiv220703561R,2022Univ....8...94C,2023arXiv230111814V}). Neutrino flavor conversion is a representative quantum feature, and various types of neutrino flavor conversions associated with neutrino self-interactions occur in CCSNe \cite{2021PhRvD.103f3033A,2021PhRvD.104h3025N,2023arXiv231111272A} and BNSMs \cite{2017PhRvD..95j3007W,2023PhRvD.108h3002X}. On the other hand, the length scale of flavor conversions is extremely smaller than the astrophysical size, making the first-principles simulations intractable. Although neutrino-radiation-hydrodynamic simulations have matured significantly, one should keep in mind that large uncertainties still remain concerning impacts of neutrino flavor conversions even in the current state-of-the-art numerical simulations. Since neutrino-matter interactions depend on neutrino flavors, flavor conversions change the feedback to the fluid dynamics \cite{2023PhRvL.130u1401N,2023PhRvD.107j3034E,2023PhRvL.131f1401E} and also nucleosynthesis \cite{2020ApJ...900..144X,2020PhRvD.102j3015G,2021PhRvL.126y1101L,2022PhRvD.106j3003F,2022PhRvD.105h3024J,2023MNRAS.519.2623F}. We also note that the dynamics of flavor conversion and its asymptotic behavior hinge on global advection of neutrinos \cite{2022PhRvL.129z1101N,2023PhRvL.130u1401N,2023PhRvD.107h3016X,2023PhRvD.108j3014N,2023PhRvD.107f3025S}, exhibiting that global neutrino-radiation-hydrodynamic simulations with incorporating effects of flavor conversions are mandatory to study the astrophysical consequence of flavor conversions.

There are respectable previous work that incorporate effects of neutrino flavor conversion in global neutrino-radiation-hydrodynamic simulations in CCSNe and BNSMs \cite{2021PhRvL.126y1101L,2022PhRvD.105h3024J,2022PhRvD.106j3003F,2023PhRvD.107j3034E,2023PhRvL.131f1401E}. Although the details vary, they commonly add a neutrino-mixing prescription on top of their classical neutrino transport schemes, in which they shuffle neutrino flavors one way or another. It should be noted that all mixing schemes employ rather phenomenological treatments and, hence, these results need to be considered provisional. This is mainly because the current implementation of flavor conversion in their codes are rather schematic, which does not have the ability to draw robust conclusions about impacts of flavor conversions. Improving their neutrino mixing schemes is obviously needed, but it is very hard along with proposed approaches. More importantly, it is not clear how we can give feedback from the results of fully quantum kinetic neutrinos to these phenomenological models. This paper is meant to address this issue and to provide a new way to fill the gap between phenomenological and first-principle simulations.

In this paper, we propose another coarse-grained neutrino transport approach: subgrid-scale modeling for neutrino flavor conversions. We distinguish our method from other phenomenological approaches, since the method is designed so as to reproduce the spatially- and time-averaged features of neutrino flavor conversions obtained from quantum kinetic neutrino simulations. The noticeable advantage in our subgrid model is having a refinable formulation for dynamics of flavor conversions by various ways including analytic methods \cite{2023PhRvD.107f3033N,2023PhRvD.107j3022Z,2023PhRvD.107l3021Z,2023PhRvD.108f3003X} and artificial intelligence (AI) techniques \cite{2023arXiv231115656A}. In this paper, we also demonstrate classical neutrino transport simulations with the subgrid model, in which we employ a simple but physically motivated subgrid model for flavor conversions.

This paper is organized as follows. In Sec.~\ref{sec:basiceq}, we start with explaining the philosophy of our proposed method. We then provide the quantum kinetic equation with our subgrid model. We also provide its two-moment formalism in Sec.~\ref{sec:twomoment}. These transport equations are written in terms of the 3+1 general relativistic formulation, which would be helpful for those who work on CCSN and BNSM simulations. After we discuss some details of the method in Sec.~\ref{sec:tscale}, we highlight novelties of our subgrid model by comparing to other phenomenological approaches in Sec.~\ref{sec:otherPheno}. In Sec.~\ref{sec:misc}, we also discuss the relevance to another coarse-grained approach: miscidynamics \cite{2023arXiv230614982J}. As shall be shown in the section, this formulation is closely associated with our formulation, indicating that both approaches are complementary to each other. To show the capability of our subgrid model, we demonstrate numerical simulations by using both quantum kinetic neutrino transport and classical one with subgrid model, paying attention to fast neutrino-flavor conversion (FFC) in Sec.~\ref{sec:demo}. By comparing their results, we can learn the source of error in the subgrid model. We then discuss strategies how to improve them based on studies of quantum kinetic neutrino transport. Finally, we summarize our work in Sec.~\ref{sec:summary}. Otherwise stated, we work in the unit with $c = \hbar = 1$, where $c$ and $\hbar$ are the speed of the light and the reduced Planck constant, respectively. In this paper, we will describe all equations with the metric signature of $- + + +$.

\section{Basic equation for neutrino transport with BGK subgrid modeling}\label{sec:basiceq}
It has been discussed that neutrino flavor conversions have quasi-steady and asymptotic behaviors in the non-linear phase \cite{2021PhRvD.104j3003W,2021PhRvD.104j3023R,2022PhRvD.106j3039B,2022PhRvD.106d3011R,2023PhRvD.107j3022Z,2023PhRvD.107l3021Z,2023PhRvD.108f3003X,2023arXiv231113842Z} or quasi-periodic properties represented as pendulum motions in flavor space \cite{2006PhRvD..74j5010H,2020PhRvD.101d3009J,2022PhRvL.128l1102P,2023PhRvD.107l3024F,2023arXiv231210340J,2023PhRvD.107d3024F,2023arXiv231207612F}. We are interested in the time- and spatially averaged states in the late non-linear phase, since it is unlikely that fine structures with short-time or small-length variations affect astrophysical consequences. Motivated by these studies, we assume that flavor conversions make the radiation field settle into an asymptotic state, and the asymptotic density matrix of neutrinos is denoted by $f^a$.

In general, the non-linear evolution of flavor conversions is very complex, and the detail hinges on flavor instabilities, neutrino-matter interactions, and global geometries of radiation fields. On the other hand, there is always a characteristic timescale of flavor conversions or associated flavor instabilities, which is denoted by $\tau_a$ in the following discussion. We note that the timescale depends on neutrino energy, angle, and neutrino flavor. $\tau_a$ also provides a rough estimation of timescale that the density matrix of neutrinos settles into $f^a$.

The quantum kinetic equation (QKE) for neutrino transport can be written as
\begin{equation}
\begin{aligned}
p^{\mu} \frac{\partial f}{\partial x^{\mu}}
+ \frac{dp^{i}}{d\tau} \frac{\partial f}{\partial p^{i}}
= - p^{\mu} u_{\mu} S
  + i p^{\mu} n_{\mu} [H,f],
\end{aligned}
\label{eq:QKEorig}
\end{equation}
where $f$ denotes the density matrix of neutrinos. In the expression, $p^{\mu}$, $x^{\mu}$, and $\tau$ denote neutrino four momentum, spacetime coordinates, and affine parameter for trajectories of neutrinos, respectively. $u^{\mu}$, $n^{\nu}$, $S$, and $H$ appearing in the right hand side of Eq.~\ref{eq:QKEorig} represent four-velocity of fluid, the unit vector normal to the spatial hypersurface in four dimensional spacetimes, collision term, and neutrino oscillation Hamiltonian, respectively. Below, we approximate Eq.~\ref{eq:QKEorig} by using $f^a$ and $\tau_a$.

Our subgrid model is developed based on an assumption that the neutrino distributions are relaxed to $f^a$ by flavor conversions in the timescale of $\tau_a$. This corresponds to a relaxation-time approximation proposed by Bhatnagar–Gross–Krook (BGK) \cite{1954PhRv...94..511B}, in which they use the approximation to collision term in Boltzmann equation for gas dynamics. In our BGK subgrid model, we apply the model to the neutrino oscillation Hamiltonian (the second term in the right hand side of Eq.~\ref{eq:QKEorig}),
\begin{equation}
\begin{aligned}
p^{\mu} \frac{\partial f}{\partial x^{\mu}}
+ \frac{dp^{i}}{d\tau} \frac{\partial f}{\partial p^{i}}
= - p^{\mu} u_{\mu} S
  + p^{\mu} n_{\mu} \frac{1}{\tau_{a}} ( f - f^{a} ).
\end{aligned}
\label{eq:BGKQKE}
\end{equation}
We note that the relaxation-time ($\tau_a$) is measured in laboratory (or $n$) frame, but it can be changed based on the fluid rest frame (see also \cite{2022PhRvD.106f3011N}), which may be useful for the frequently used two-moment formalism for neutrino transport (see Sec.~\ref{sec:twomoment}). It should also be noted that $f^a$ and $\tau_a$ are determined from $f$ at each time step, implying that they are time-dependent quantities.

It should be mentioned that the BGK subgrid model (or relaxation-time approximation) is applicable to any systems for which there is an equilibrium (or asymptotic) state. As shown in \cite{2024arXiv240208896J}, neutrino flavor conversion is ergodic (at least approximately), exhibiting that the dynamical feature is similar to thermodynamics. The equilibration occurs because it's the most probable (entropy-maximizing) outcome (see also \cite{2023arXiv230614982J}).

It is worth noting that a similar approximation was used to obtain a temporally coarse-grained quantum kinetic equation for the production of sterile neutrinos (see Eqs. 4 and 5 of \cite{2019PhRvD.100h3536J}). There it was proposed that the entire right-hand side, including both oscillation and collision terms, be treated using a BGK approximation. This ansatz showed excellent agreement with numerical results. Here we adapt the relaxation-time approximation to the context of collective neutrino oscillations by proposing that it can be applied to oscillations alone, with subgrid relaxation being caused by collective modes rather than collisions.

From a practical point of view, we also provide a conservative form of Eq.~\ref{eq:BGKQKE}, which is used for numerical simulations for both Boltzmann- and quantum kinetic neutrino transport (see, e.g., \cite{2017ApJS..229...42N,2022PhRvD.106f3011N}). Following \cite{2014PhRvD..89h4073S}, we can rewrite the transport equation as,
\begin{equation}
  \begin{split}
&\frac{1}{\sqrt{-g}} \left. \frac{\partial}{\partial x^{\alpha}} \right|_{q_{i}}
\Biggl[  \Bigl( n^{\alpha} + \sum^{3}_{i=1} \ell_{(i)} e^{\alpha}_{(i)} \Bigr) \sqrt{-g} f   \Biggr] \\
& - \frac{1}{\varepsilon^2} \frac{\partial}{\partial \varepsilon}( \varepsilon^3 f \omega_{(0)}  )
+ \frac{1}{\sin\theta_{\nu}} \frac{\partial}{\partial \theta_{\nu}}
( \sin\theta_{\nu} f \omega_{(\theta_{\nu})} ) \\
& + \frac{1}{ \sin^2 \theta_{\nu}} \frac{\partial}{\partial \phi_{\nu}} (f \omega_{(\phi_{\nu})}) 
= D S - \frac{1}{\tau_{a}} ( f - f^{a} ).
  \end{split}
\label{eq:conformQKE}
\end{equation}
In the expression, $\varepsilon$ and $g$ are the neutrino energy measured from $e^{\alpha}_{(0)}=n^{\alpha}$ observer, i.e., $\varepsilon \equiv - p_{\alpha} n^{\alpha}$, and the determinant of the four-dimensional metric, respectively. $e^{\alpha}_{(i)} (i = 1, 2, 3)$ denote a set of the (spatial) tetrad bases normal to $n$. $\theta_{\nu}$ and $\phi_{\nu}$ denote the neutrino flight direction in the laboratory (or $n$) frame. These angles are measured from $e^{\alpha}_{(1)}$, and the three coefficients of $\ell_{i}$ represents the directional cosines, which can be expressed as,
\begin{equation}
  \begin{split}
&\ell_{(1)} = \cos \hspace{0.5mm} \theta_{\nu}, \\
&\ell_{(2)} = \sin \hspace{0.5mm} \theta_{\nu}   \cos \hspace{0.5mm} \phi_{\nu}, \\
&\ell_{(3)} = \sin \hspace{0.5mm} \theta_{\nu}   \sin \hspace{0.5mm} \phi_{\nu}.
  \end{split}
\label{eq:el}
\end{equation}
$D$ in the right hand side of Eq.~\ref{eq:conformQKE} represents the effective Doppler factor, which is defined as $D \equiv \nu / \varepsilon$ with $\nu \equiv - p^{\mu} u_{\mu}$, while $\nu$ denotes the neutrino energy measured in the fluid rest frame. $\omega_{(0)}, \omega_{(\theta_{\nu})}, \omega_{(\phi_{\nu})}$ appearing in the left hand side of Eq.~\ref{eq:conformQKE} can be written as,
\begin{equation}
  \begin{split}
& \omega_{(0)} \equiv \varepsilon^{-2} p^{\alpha} p_{\beta} \nabla_{\alpha} n^{\beta}, \\
& \omega_{(\theta_{\nu})} \equiv \sum^{3}_{i=1} \omega_{i} \frac{ \partial \ell_{(i)} }{\partial \theta_{\nu} }, \\
& \omega_{(\phi_{\nu})} \equiv \sum^{3}_{i=2} \omega_{i} \frac{ \partial \ell_{(i)} }{\partial \phi_{\nu} }, \\
&\omega_{i} \equiv \varepsilon^{-2} p^{\alpha} p_{\beta} \nabla_{\alpha} e^{\beta}_{(i)}.
  \end{split}
\label{eq:Omega}
\end{equation}
Spherical polar coordinate is often employed in multi-angle neutrino transport codes (see e.g., \cite{2012ApJS..199...17S,2014ApJS..214...16N,2022PhRvD.106f3011N}). We, hence, chose a set of tetrad basis, $\mbox{\boldmath $e$}_{(i)}$ as,
\begin{equation}
  \begin{split}
& e^{\alpha}_{(1)} = (0, \gamma^{-1/2}_{rr}, 0, 0 ) \\
& e^{\alpha}_{(2)} = \Biggl(0, -\frac{\gamma^{-1/2}_{r \theta}}{\sqrt{\gamma_{rr} (\gamma_{rr} \gamma_{\theta \theta} - \gamma^2_{r \theta})}}, \sqrt{ \frac{\gamma_{rr}}{ \gamma_{rr} \gamma_{\theta \theta} - \gamma^2_{r \theta} } }, 0 \Biggr) \\
& e^{\alpha}_{(3)} = \Biggl(0, \frac{\gamma^{r \phi}}{\sqrt{\gamma^{\phi \phi}}} , \frac{\gamma^{\theta \phi}}{\sqrt{\gamma^{\phi \phi}}}, \sqrt{\gamma^{\phi \phi}} \Biggr),
  \end{split}
\label{eq:polartetrad}
\end{equation}
where $\gamma^{\alpha \beta} \equiv g^{\alpha \beta} + n^{\alpha} n^{\beta}$.

One thing we do notice here is that Eq.~\ref{eq:BGKQKE} (or \ref{eq:conformQKE}) corresponds to a classical transport equation, if we neglect the off-diagonal elements. Since the main purpose of this study is to provide a subgrid model of neutrino flavor conversion for classical neutrino transport schemes, we limit our discussion only for the classical transport with BGK subgrid model. One should keep in mind that the subgrid model can be applied to neutrino quantum kinetics, and appropriate modeling of off-diagonal components would increase the physical fidelity of subgrid model. This is an intriguing possibility and deserves further investigations, although we postpone the study to future work.

Below, let us consider how to determine diagonal components of $f^a$. It is well known that the lepton number of neutrinos/antineutrinos does not change during flavor conversions. This indicates that we can characterize $f^a$ via survival probability of neutrinos ($\eta$), while it depends on neutrino energy and flight angle, in general. Following the prescriptions in \cite{2000PhRvD..62c3007D,2021MNRAS.500..696N,2021MNRAS.500..319N,2023PhRvD.107f3033N}, we can write $f^a$ in terms of $f$ as,
\begin{equation}
 \begin{split}
&f_{e}^{a}  = \eta  f_{e}  + \left(1- \eta  \right) f_{x} , \\
&f_{x}^{a}  = \frac{1}{2}\left(1- \eta  \right) f_{e}  + \frac{1}{2} \left(1+\eta  \right)  f_{x},
 \end{split}
\label{eq:trapro_three}
\end{equation}
where $f_{e}$ and $f_{x}$ represent distribution functions (or diagonal elements of density matrix) for electron-type and heavy-leptonic type neutrinos, respectively. We note that $\mu$ and $\tau$ neutrinos are assumed to be the same in Eq.~\ref{eq:trapro_three}, which is a reasonable assumption for CCSNe and BNSMs. However, they are quantitatively different from each other, in particular for high energy neutrinos (see, e.g., \cite{2021MNRAS.502...89N}), due to high-order corrections in neutrino-matter interactions (e.g., weak-magnetism \cite{2002PhRvD..65d3001H}). We also note that, if on-shell muons appear \cite{2017PhRvL.119x2702B,2020PhRvD.102l3001F,2020PhRvD.102b3037G}, we should distinguish $\mu$- and $\tau$ neutrinos. We can deal with these cases by introducing another parameter to represent neutrino mixing. For antineutrinos, we can use the same form as Eq.~\ref{eq:trapro_three} but replacing $f$ and $\eta$ to $\bar{f}$ and $\bar{\eta}$, respectively.

There are two important remarks about our BGK subgrid model. First, $f^a$ (or $\eta$) hinges on flavor instabilities, and it should be determined (or calibrated) based on neutrino quantum kinetics. It is important to note that the results from analytic studies and local simulations of flavor conversions can be directly used to determine it. In Sec.~\ref{sec:demo}, we demonstrate such simulations for FFC. Second, if the system contains multiple flavor instabilities, we can handle the problem with multiple BGK terms. More specifically, the second term in right hand side of Eq.~\ref{eq:BGKQKE} can be rewritten as,
\begin{equation}
\begin{aligned}
p^{\mu} n_{\mu} \frac{1}{\tau_{a}} ( f - f^{a} ) \rightarrow
p^{\mu} n_{\mu} \sum_{i=1}^{N}
\frac{1}{\tau_{a_i}} ( f - f^{a_i} )
\end{aligned}
\label{eq:multipleBGK}
\end{equation}
where the index $i$ distinguish flavor instabilities among $N$ modes. 
As shown in Eq.~\ref{eq:multipleBGK}, the contribution of each term is characterized by $\tau_{a_i}$ and $f - f^{a_i}$, which guarantees that flavor conversion with shorter relaxation time and large difference between $f$ and $f^{a_i}$ dominate the system. This prescription may be important for realistic CCSN and BNSM models, since FFC and collisional flavor instabilities (CFI) may occur simultaneously (see, e.g., \cite{2023arXiv231111272A}) at the same position. The extension by Eq.~\ref{eq:multipleBGK} allows us to study the situation where multiple flavor instabilities are competing to each other.

Before we discuss how to estimate $\tau_a$ in Sec.~\ref{sec:tscale}, let us describe the two-moment transport formalism for our subgrid model in the next section. This is helpful for those who use the moment formalism for numerical modeling of CCSNe and BNSMs.

\section{Two-moment formalism}\label{sec:twomoment}
Moment formalism of radiation transport has, in principle, the ability to describe full neutrino kinetics with equivalent level of Boltzmann (or fully quantum kinetic) neutrino transport. In practice, however, the moment formalism results in infinite hierarchy of coupled equations, indicating that we need to truncate the hierarchy of moments at a certain rank. The currently most popular approach in neutrino transport simulations is two-moment formalism \cite{2011PThPh.125.1255S,2013PhRvD..87j3004C,2015PhRvD..91l4021F,2015MNRAS.453.3386J,2016ApJS..222...20K,2018ApJ...854...63O,2019ApJS..241....7S,2020MNRAS.495.2285W,2020LRCA....6....4M,2021ApJS..253...52L,2022MNRAS.512.1499R,2023ApJS..267...38C}, in which the zeroth and first angular moments correspond to fundamental variables. We determine their time evolution and spatial distributions by solving their coupling equations, while higher-rank moments are complemented by closure relations. It is worth noting that the moment formalism is also used for the study of neutrino flavor conversions
\cite{2020PhRvD.101d3009J,2020arXiv200909024J,2022PhRvD.105l3036M,2023PhLB..84638210G,2023arXiv231111968F}. In this section, we provide an explicit description of two-moment formalism with BGK subgrid model.

Following the convention of \cite{2011PThPh.125.1255S}, we decompose the neutrino four momentum ($p^{\alpha}$) into $u^{\alpha}$ and its orthogonal normal vector ($\ell^{\alpha}$) as,
\begin{equation}
\begin{aligned}
p^{\alpha} = \nu (u^{\alpha} + \ell^{\alpha}),
\end{aligned}
\label{eq:fourmomentumdecom}
\end{equation}
while the conditions of $\ell^{\alpha} u_{\alpha}=0$ and $\ell^{\alpha} \ell_{\alpha} = 1$ are satisfied. The unprojected second- and third rank moments of neutrinos are defined as (see also \cite{1981MNRAS.194..439T}),
\begin{equation}
  \begin{split}
M^{\alpha \beta} & \equiv \nu^3 \int f (u^{\alpha} + \ell^{\alpha}) (u^{\beta} + \ell^{\beta}) d \Omega , \\
M^{\alpha \beta \gamma} & \equiv \nu^3 \int f (u^{\alpha} + \ell^{\alpha}) (u^{\beta} + \ell^{\beta}) (u^{\gamma} + \ell^{\gamma}) d \Omega , \\
  \end{split}
\label{eq:enemomtensor}
\end{equation}
where $\Omega$ denotes the solid angle of neutrino momentum space defined in the fluid-rest frame. It should be mentioned that the integral of $M^{\alpha \beta}$ over the neutrino energy ($\int M^{\alpha \beta} d \nu$) corresponds to the energy-momentum tensor of neutrinos. We also define the zeroth and first angular moments defined in the fluid-rest frame as,
\begin{equation}
  \begin{split}
 J & \equiv \nu^{3} \int f d \Omega , \\
 H^{\alpha} & \equiv \nu^{3} \int \ell^{\alpha} f d \Omega, \\ 
 L^{\alpha \beta} & \equiv \nu^{3} \int \ell^{\alpha} \ell^{\beta} f d \Omega, \\ 
 N^{\alpha \beta \gamma} & \equiv \nu^{3} \int \ell^{\alpha} \ell^{\beta} \ell^{\gamma} f d \Omega, \\ 
  \end{split}
\label{eq:Omega}
\end{equation}

By using these variables, the basic equation for the two-moment formalism with BGK subgrid model can be written as (see also Eq.~\ref{eq:BGKQKE}),
\begin{equation}
  \begin{split}
\nabla_{\beta} M^{\alpha \beta} - \frac{\partial}{\partial \nu} 
(\nu M^{\alpha \beta \gamma} \nabla_{\gamma} u_{\beta})
= S^{\alpha} - W^{\alpha}
  \end{split}
\label{eq:twomom_sourceEq}
\end{equation}
where
\begin{equation}
  \begin{split}
S^{\alpha} & \equiv \nu^3 \int S (u^{\alpha} + \ell^{\alpha})  d \Omega , \\
W^{\alpha} & \equiv \frac{1}{\tau_a^{\rm fl}} \nu^3 \int (f-f^a) (u^{\alpha} + \ell^{\alpha}) d \Omega, \\
  \end{split}
\label{eq:defSandW}
\end{equation}
where $\tau_a^{\rm fl} \equiv D \hspace{0.5mm} \tau_a$.
Eq.~\ref{eq:twomom_sourceEq} indicates that the BGK subgrid model can be implemented simply by replacing $S^{\alpha} \to S^{\alpha} - W^{\alpha}$ from the original two-moment formalism. $W^{\alpha}$ can be expressed similar form as emission-absorption process of collision term, which can be written as,
\begin{equation}
\begin{aligned}
W^{\alpha} = \frac{1}{\tau_a^{\rm fl}} 
\biggl(
( J - J^a )u^{\alpha} + ( H^{\alpha} - H^{\alpha a} )
\biggr).
\end{aligned}
\label{eq:WbyJandH}
\end{equation}
We, hence, need to determine $\tau_a$, $J^{a}$, and $H^{\alpha a}$ to implement the BGK model.

$J^{a}$ and $H^{\alpha a}$ can be obtained by taking angular integrals of Eq.~\ref{eq:trapro_three}, and it looks that the process is straightforward. However, $\eta$ depends on $\Omega$ in general, indicating that we need higher-rank angular moments to evaluate them. Below, we provide an approximate prescription to address this issue.

We start with expanding the angular dependence of $\eta$ by $\ell_{\alpha}$ as,
\begin{equation}
\begin{aligned}
\eta = \eta_{0} + \eta_{1}^{\alpha} \ell_{\alpha} + \eta_{2}^{\alpha \beta} \ell_{\alpha} \ell_{\beta} + ..... ,
\end{aligned}
\label{eq:eta_angexpand}
\end{equation}
where the coefficients ($\eta_{i}$) do not depend on $\Omega$. By using the expression, $J^{a}$ and $H^{\alpha a}$ can be written as,
\begin{equation}
 \begin{split}
J_{e}^{a} = & \hspace{0.5mm} J_x + \eta_{0} ( J_e - J_x ) + \eta_{1}^{\alpha} ( H_{e \alpha} - H_{x \alpha} ) \\
& + \eta_{2}^{\alpha \beta} ( L_{e \alpha \beta} - L_{x \alpha \beta} ) + ....  \\
H_{e}^{\alpha a} = & \hspace{0.5mm} H_x^{\alpha} + \eta_{0} ( H_e^{\alpha} - H_x^{\alpha} ) + \eta_{1}^{\beta} ( L_{e \beta}^{\alpha} - L_{x \beta}^{\alpha} ) \\
& + \eta_{2}^{\beta \gamma} ( N_{e \beta \gamma}^{\alpha} - N_{x \beta \gamma}^{\alpha} ) + ....  \\
J_{x}^{a} = & \hspace{0.5mm} \frac{1}{2}(J_e + J_x) - \frac{\eta_{0}}{2} ( J_e - J_x ) - \frac{\eta_{1}^{\alpha}}{2} ( H_{e \alpha} - H_{x \alpha} ) \\
& - \frac{\eta_{2}^{\alpha \beta}}{2} ( L_{e \alpha \beta} - L_{x \alpha \beta} ) + .... \\
H_{x}^{\alpha a} = & \hspace{0.5mm} \frac{1}{2}(H_{e}^{\alpha} + H_{x}^{\alpha}) - \frac{\eta_{0}}{2} ( H_{e}^{\alpha} - H_{x}^{\alpha} ) - \frac{\eta_{1}^{\beta}}{2} ( L_{e \beta}^{\alpha} - L_{x \beta}^{\alpha} ) \\
& - \frac{\eta_{2}^{\beta \gamma}}{2} ( N_{e \beta \gamma}^{\alpha} - N_{x \beta \gamma}^{\alpha} ) + ....
 \end{split}
\label{eq:JaHa_fullAng}
\end{equation}
This method guarantees that flavor-integrated angular moments are conserved regardless of $\eta_{i}$, even if we truncate their angular moments at any order.

Eq.~\ref{eq:JaHa_fullAng} exhibits that the accuracy of determining $J^{a}$ and $H^{\alpha a}$ hinges on how well we can determine coefficients $\eta_{i}$. In two-moment neutrino transport code, the maximum-entropy completion \cite{2021PhRvD.103l3012J,2022PhRvD.106h3005R} (or a fitting method proposed in \cite{2021PhRvD.103l3025N}, which can be used only for CCSNe, though) may be useful to obtain physically reasonable solutions. A noticeable feature in these methods is that we approximately reconstruct full angular distributions of neutrinos from their zeroth and first angular moments. This suggests that the angular dependence of $\eta$ can also be determined by a similar way as multi-angle neutrino transport (see in Sec.~\ref{subsec:numsetBGKSM} for more details).

Neglecting energy-dependence and anisotropic components in $\eta$, i.e., $\eta(\nu,\Omega) = \eta_0$, corresponds to the simplest case, but it would be a reasonable approximation for CFI. Since the CFI becomes important in regions where neutrinos and matters are tightly coupled, neutrinos are nearly isotropic in momentum space \cite{2023arXiv231005050L,2023arXiv231111272A}. We also note that the so-called isotropy-preserving branch in $k=0$ mode provides the maximum growth rate of the instability \cite{2023PhRvD.107l3011L}, lending confidence to diminishing angular dependence in $\eta$. Regarding the energy dependence, on the other hand, the authors in \cite{2023PhRvD.107l3011L} found that the growth rate of CFI can be well approximated by the monochromatic energy treatment with averaged-energy collision rates. We also note that flavor swap is accompanied by resonance-like CFI, but the dynamics does not depend on neutrino energy \citep{2023arXiv230902619K}, suggesting that the energy dependence is not important in these cases. 

The condition, $\eta(\nu,\Omega) = \eta_0$, corresponds to the simplest case for our BGK model but it would be useful to explore qualitative trends for impacts of flavor conversions on CCSN and BNSM, as studied with phenomenological approaches. It should be emphasized that our subgrid model takes into account the relaxation-time scale, indicating that the interaction between neutrino advection, neutrino-matter interaction, and flavor conversions would be more appropriately handled than other phenomenological ones. It seems that $\eta_0=1/3$ and $0$ are two interesting cases, which correspond to flavor equipartition and flavor swap, respectively.

\section{Estimation for $\tau_a$}\label{sec:tscale}
The vigor of flavor conversion can not be measured only by $f^a$. Even if the asymptotic distribution is very different from the original non-mixing state, the flavor conversion can not be completed if the relaxation-time is very long. This exhibits that the determination of $\tau_a$ is also important task to increase the accuracy of our subgrid model. 

Linear stability analysis can offer the growth rate of flavor conversion, which would be the most accurate determination of $\tau_a$. However, the growth rate can be obtained by solving the dispersion relation (see, e.g., \cite{2017PhRvL.118b1101I,2018JCAP...12..019A}), which is a computationally expensive task. We also note that, in the stability analysis, full energy- and angular dependent information of neutrinos in momentum space are required in general, but they can be obtained only by solving multi-angle and multi-energy neutrino transport, indicating that these information are not available for approximate neutrino transport. We, hence, need alternative approaches for the estimation of $\tau_a$ to suit our need.

We can utilize some approximate approaches of the stability analysis, that have been proposed in the literature. For FFC, a simple formula was provided in \cite{2019ApJ...886..139N,2020PhRvR...2a2046M}. In this method, we can approximately estimate $\tau_a$ as,
\begin{equation}
\begin{aligned}
\tau_a \sim 2 \pi \left| \left(\int_{G_{v}>0} d\Gamma G_{v}\right)\left(\int_{G_{v}<0} d\Gamma G_{v}\right)   \right|^{-1/2}, \label{eq:approxiGrowth_NaMo}
\end{aligned}
\end{equation}
where
\begin{equation}
\begin{split}
d \Gamma_{v} &\equiv \frac{1}{4 \pi} d(\cos{\theta_{\nu}}) d \phi_{\nu} \\
G_{v} &\equiv \frac{1}{2 \pi^2} \int \biggl( (f_{e} - \bar{f}_{e}) - (f_{x} - \bar{f}_{x}) \biggr) \varepsilon^2 d \varepsilon.
\label{eq:Gamma_and_Gv}
\end{split}
\end{equation}
In Sec.~\ref{sec:demo}, we demonstrate neutrino transport simulations for FFC by using Eq.~\ref{eq:Gamma_and_Gv}.

It is also note-worthy that Eq.~\ref{eq:approxiGrowth_NaMo} is applicable for two-moment method by using the maximum-entropy completion \cite{2021PhRvD.103l3012J,2022PhRvD.106h3005R} or a fitting method \cite{2021PhRvD.103l3025N}, since they can approximately retrieve $f$ from the zeroth and first angular moments. It would also be useful to employ other methods as in \cite{2018PhRvD..98j3001D,2020JCAP...05..027A,2021PhRvD.104f3014N,2021PhRvD.103l3012J,2022PhRvD.106h3005R,2023PhRvD.107l3011L,2023arXiv231111968F}, which allows us to evaluate the growth rate of flavor conversions directly from low angular moments of neutrinos. For CFI, the growth rate can also be estimated analytically \cite{2021arXiv210411369J,2023PhRvD.108h3002X,2023PhRvD.107l3011L}, which is also useful for our subgrid model. We can select them depending on the problem and the purpose of study. Another remark here is that machine-learning techniques potentially provide accurate estimations of $\eta$ and $\tau_a$ without significant computational burden (see, e.g., \cite{2023PhRvD.107j3006A,2023arXiv231003807A,2023arXiv231115656A})

\section{Comparing to other phenomenological models}\label{sec:otherPheno}
It would be worthwhile to highlight differences of our sub-grid model from other phenomenological methods implemented in some neutrino-radiation-hydrodynamic codes. The study by \cite{2021PhRvL.126y1101L} corresponds to a pioneer work for BNSM simulations with a phenomenological model of FFC, in which effects of FFC are incorporated by a parametric prescription. In their method, occurrences of FFC are identified based on $k=0$ mode stability analysis. They shuffle neutrinos between $\nu_e$, $\nu_{\mu}$, and $\nu_{\tau}$ to be flavor equipartition, if the time scale of flavor conversion is shorter than the critical one (which was assumed to be $10^{-7}$s). This indicates that their prescription of flavor conversion can be reproduced in our sub-grid model by setting $\eta=1/3$ and $\tau_a \to 0$, if the growth time scale is shorter than $10^{-7}$s (otherwise $\tau_a$ is set to be infinity).

In \cite{2022PhRvD.105h3024J}, they also carried out BNSM simulations by a similar approach as \cite{2021PhRvL.126y1101L}, but they study impacts of FFCs on BNSM dynamics by considering three types of neutrino mixing schemes. Essentially, the degree of neutrino mixing varies among schemes, while the detection criterion for occurrences of FFC is common, in which they determine FFCs only by energy-averaged flux factor of $\bar{\nu}_e$. They also assumed that flavor conversions occur instantaneously (i.e., $\tau_a \to 0$ in our BGK subgrid model). This approach can also be reproduced by our subgrid model.

The similar study for FFCs in BNSM has also been made by \cite{2022PhRvD.106j3003F}. Different from \cite{2021PhRvL.126y1101L,2022PhRvD.105h3024J}, they employed a so-called leakage scheme for neutrino transport. In their method, the neutrino transport scheme is left as the original, but they changed the estimation of neutrino luminosity by taking into account FFCs, which corresponds to a key ingredient in their scheme to give a feedback of neutrinos to fluid dynamics and ejecta compositions. They determine asymptotic neutrino luminosities by varying parameters (including cases with flavor equipartition), while they also employ neutrino opacities to determine the degree of mixing. In their approach, flavor conversions are suppressed in optically thick region, whereas they occur in optically thin one. Since this phenomenological model is developed based on a different philosophy from ours, our subgrid model can not reproduce their model. Nevertheless, it is interesting to compare our subgrid model to their phenomenological model in CCSN and BNSM simulations.

Impacts of FFC on CCSN dynamics have also been studied by another phenomenological approach in \cite{2023PhRvD.107j3034E,2023PhRvL.131f1401E}. In their method, the number of independent neutrino flavors are three: $\nu_e$, $\bar{\nu}_e$, and $\nu_x$, while they shuffle them so as to guarantee the neutrino lepton number conversion. They employ matter density to determine occurrences of FFC, in which there is a threshold density that flavor conversions occur. In the region where the matter density is lower than the threshold, they assume that neutrino flavor conversions occur instantaneously. They also assume that neutrinos are in flavor equilibrium, but $\nu_x$ and $\bar{\nu}_x$ are assumed to be identical after the conversion is completed. As such, this phenomenological model is developed based on a very different approach from our subgrid one.

One of the interesting applications for our subgrid model is to assess the capability of each phenomenological model. The assessment has been impossible thus far by direct numerical simulations of quantum kinetic neutrino transport due to extremely high computational cost, but it is feasible by using our subgrid model. This study would also help us to improve each phenomenological model.

\section{Comparing to miscidynamics \label{sec:misc}}

The coarse-grained subgrid model is compatible with the proposal to approximate neutrino quantum kinetics using neutrino quantum thermodynamics \cite{2023arXiv230614982J}. Taking $\tau_a \rightarrow 0$ in Eq.~\ref{eq:BGKQKE} results in
\begin{equation}
\begin{aligned}
p^{\mu} \frac{\partial f^a}{\partial x^{\mu}}
+ \frac{dp^{i}}{d\tau} \frac{\partial f^a}{\partial p^{i}}
= - p^{\mu} u_{\mu} S^a,
\end{aligned}
\end{equation}
where $S^a$ is $S$ evaluated using $f = f^a$. This equation is equivalent to the miscidynamic transport equation written down in Ref.~\cite{2023arXiv230614982J} if $f^a$ is equated to $\rho^\textrm{eq}$ in that paper.

Miscidynamics refers to coarse-grained neutrino transport based on the concept of local mixing equilibrium. Our subgrid model does not necessarily assume that $f^a$ is an equilibrium state in a thermodynamic sense. If we do assume this, however, then taking the limit of short relaxation-time $\tau_a$ is a means of imposing local mixing equilibrium. The thermodynamic input then enters through the determination of $f^a$.

If $\tau_a \rightarrow 0$, neutrino flavor instantaneously equilibrates, and therefore it should never depart from equilibrium in the first place. This is the idea behind the adiabatic proposal of Ref.~\cite{2023arXiv230614982J}. Accepting this logic, it is then possible to determine $f^a$ using the assumption of adiabaticity and the requirements of self-consistency. Adiabaticity relates $f$ to the Hamiltonian, but the Hamiltonian is itself a function of $f$ through neutrino--neutrino forward scattering, hence the need for self-consistency. In the more straightforward case of MSW flavor conversion without neutrino self-interactions, self-consistency is not required and $f^a$ is simply determined by vacuum oscillations and neutrino--matter forward scattering.

Finite equilibration rates entail some amount of entropy production. Formulating diabatic miscidynamics---in contrast with the adiabatic version described above---would require a consideration of how subgrid degrees of freedom in the neutrino flavor field respond to grid-level changes driven by the derivative and collisional terms in Eq.~\ref{eq:BGKQKE}. Generally speaking, if the microscopic constituents respond very quickly, then the macroscopic system moves between equilibria with minimal entropy production. Equations supplementing miscidynamics with diabatic terms have not yet been worked out. In their absence, a relaxation-time $\tau_a$ is a simple and plausible approximation of diabaticity.

One subtlety in using our BGK subgrid model for diabatic miscidynamics is that $f^a$ changes under diabatic evolution. The system heats up, and mixing equilibrium is set by the system itself rather than an external environment. Because entropy production is a subgrid effect, $f^a$ can change on a subgrid timescale, which threatens the use of coarse-graining. However, a simple approximation is to adopt
\begin{equation}
    f^a \longrightarrow f^a_\infty, ~~~ \tau_a \longrightarrow \tau^\infty_a,
\end{equation}
where $f^a_\infty$ and $\tau^\infty_a$ are the $t \rightarrow \infty$ equilibrium and relaxation-time. In this approximation, neutrino flavor relaxes directly toward the ultimate equilibrium state $f^a_\infty$ rather than pursuing a time-evolving equilibrium that converges on $f^a_\infty$ at late time. The form of Eq.~\ref{eq:BGKQKE} is unchanged except for replacement of $f^a$ and $\tau_a$ by the respective asymptotic quantities.

In sum, the $\tau_a \rightarrow 0$ relaxation subgrid model can reproduce adiabatic miscidynamics. Miscidynamics can be systematically improved by calculating diabatic corrections from the statistical mechanics underlying neutrino quantum thermodynamics \cite{2023arXiv230614982J}. It appears that Eq.~\ref{eq:BGKQKE} can likewise be systematically improved by adjusting $f^a$ and $\tau_a$ to reflect these corrections.

\section{Demonstration}\label{sec:demo}
In this section, we discuss capabilities of our BGK subgrid model by carrying out local simulations of FFC in spatial one dimension (1D). Under the symmetry, neutrino angular distributions in momentum space become axisymmetric, indicating that we solve QKE for one in time, one in real space, and one in momentum space. We select this problem because analytic schemes for determining asymptotic states of FFC have been proposed in the literature \cite{2023PhRvD.107j3022Z,2023PhRvD.108f3003X}, which can be used to compute $f^a$. After we describe essential information on numerical simulations, we describe explicitly how to determine $f^a$.

\subsection{Full quantum kinetic simulations}\label{subsec:QKNT}
Here, we describe the problem under full quantum kinetic approach. Note that the results of these simulations will be used to assess simulations with BGK subgrid model; the detail will be given in Sec.~\ref{subsec:numsetBGKSM}. Quantum kinetic simulations in the present study are essentially the same as those performed in \cite{2023PhRvD.107f3033N}, in which we demonstrated 1D local simulations of FFCs in a two-flavor framework. One noticeable difference from the previous study is that we solve QKE under a three-flavor framework. Assuming spherically symmetry and no collision terms, we solve the following QKE,
\begin{equation}
  \begin{split}
& \frac{\partial \barparena{f}}{\partial t}
+ \frac{1}{r^2} \frac{\partial}{\partial r} ( r^2 \cos \theta_{\nu}  \barparena{f} )  - \frac{1}{r \sin \theta_{\nu}\
} \frac{\partial}{\partial \theta_{\nu}} ( \sin^2 \theta_{\nu} \barparena{f}) \\
&  = - i \hspace{0.5mm} [\barparena{H},\barparena{f}],
  \end{split}
\label{eq:BasEq_SpheQKE}
\end{equation}
where
\begin{equation}
\barparena{H} = \barparena{H}_{\rm vac} + \barparena{H}_{\rm mat} + \barparena{H}_{\nu \nu}, \label{eq:Hdecompose}
\end{equation}
In this expression, $f (\bar{f})$ and $H (\bar{H})$ denote the density matrix of neutrinos and the oscillation Hamiltonian for neutrinos (antineutrinos), respectively. Since we only focus on local simulations in this study, neutrino advection in $\theta_{\nu}$ direction is basically negligible. Each term of neutrino Hamiltonian can be written as,
\begin{equation}
\begin{aligned}
&\bar{H}_{\rm vac} = H^{*}_{\rm vac} , \\
&\bar{H}_{\rm mat} = - H^{*}_{\rm mat} ,\\
&\bar{H}_{\nu \nu} = - H^{*}_{\nu \nu}.
\end{aligned}
\label{eq:Hantineutrinos}
\end{equation}
Similar as \cite{2023PhRvD.107f3033N}, we ignore matter potential in Hamiltonian but their effects are effectively taken into account in vacuum potential (see below). The vacuum term is, on the other hand, included in our simulations, which has the following form,
\begin{equation}
\begin{aligned}
H_{\rm vac} = \frac{1}{2 \varepsilon} U
\begin{bmatrix}
        m^{2}_{1} & 0 & 0\\
	0 & m^{2}_{2} & 0 \\
        0 & 0 & m^{2}_{3}
    \end{bmatrix}
 U^{\dagger} ,\\
\end{aligned}
\label{eq:Hvdef}
\end{equation}
where $m_{i}^2$ and $U$ denote the neutrino squared mass for the mass eigenstate of $i$ and Pontecorvo-Maki-Nakagawa-Sakata (PMNS) matrix, respectively. Neutrino flavor conversions depend on only the difference of each squared mass of neutrino, and we set them as $\Delta m_{21}^2 = 7.42 \times 10^{-5} {\rm ev^2}$ and $\Delta m_{31}^2 = 2.510 \times 10^{-3} {\rm ev^2}$, where $\Delta m_{ij} \equiv m_{i}^2 - m_{j}^2$ in this study. We effectively include effects of matter suppression of flavor conversion by setting  the neutrino mixing angles as $10^{-6}$, which is much smaller than those constraint by experiments. It should be noted that the vacuum potential is necessary only for triggering flavor conversions, and it does not affect non-linear evolutions of FFCs. This is simply because the self-interaction potential is several orders of magnitudes higher than the vacuum one, which also guarantees that FFCs overwhelm slow modes. Throughout this test, we use a monochromatic assumption with the neutrino energy of $12$ MeV.


In setting up initial angular distributions of $\nu_e$ and $\bar{\nu}_e$, we employ the following analytic formula,
\begin{equation}
\barparena{f}_{ee} = \langle \barparena{f}_{ee}\rangle \biggl( 1 + \barparena{\beta}_{ee} ( \cos \theta_{\nu} - 0.5 )  \biggr) \hspace{4mm} \cos \theta_{\nu} \ge 0,
\label{eq:iniang_in}
\end{equation}
where $\langle f_{ee}\rangle$ corresponds to an angular-averaged distribution function for electron-type neutrinos and its bar denotes the same quantity but for antineutrinos. In this model, we vary the angular distributions of neutrinos by changing $\langle \barparena{f}_{ee}\rangle$ and $\barparena{\beta}_{ee}$. The former and latter are associated with neutrino number density and asymmetric degree of their angular distributions (see also \cite{2022PhRvL.129z1101N,2023PhRvD.107f3033N}). Similar as \cite{2023PhRvD.107f3033N}, we put a dilute neutrino gas for incoming neutrinos ($\cos \theta_{\nu} \le 0$), which do not play any roles on FFC. We also assume that there are no $\nu_{\mu}$, $\nu_{\tau}$, and their antipartners in the initial distributions. Following \cite{2023PhRvD.107f3033N}, $\langle f_{ee} \rangle$ is chosen so that the number density of $\nu_e$ becomes $10^{32} {\rm cm^{-3}}$. We determine $\langle \bar{f}_{ee}\rangle$ via a new variable, $\alpha$, which is defined as,
\begin{equation}
\alpha \equiv \frac{\langle \bar{f}_{ee}\rangle}{\langle f_{ee} \rangle} = \frac{\bar{n}_{\nu_e}}{n_{\nu_e}},
\label{eq:defalpha}
\end{equation}
where $n_{\nu_e} (\bar{n}_{\nu_e})$ denotes the number density of $\nu_e$ ($\bar{\nu}_e$). In this demonstration, we study four cases by varying $\alpha$ and $\bar{\beta}_{ee}$ while we set $\beta_{ee}=1$ for all models. The reference model corresponds to the case with $\alpha=1$ and $\bar{\beta}_{ee}=1$. We add two models by varying $\alpha$ ($\alpha=0.9$ and $1.1$), while $\bar{\beta}_{ee}$ is the same as the reference one. We test another model with $\bar{\beta}_{ee}=0.1$, while $\alpha$ is set to be the same as the reference model. It should be mentioned that the angular position for ELN crossing hinges on $\alpha$, and that $\bar{\beta}_{ee}$ dictates the depth of crossing increases (see \cite{2023PhRvD.107f3033N} for more details).

In these simulations, we focus on a spatially narrow region with $50 {\rm km} \le r \le 50 {\rm km} + 10 {\rm m}$. The radial domain and angular ($\theta_{\nu}$) direction in neutrino momentum space are covered by $N_{r}=49152$ and $N_{\theta_{\nu}}=128$ uniform grid points, respectively. We employ a Dirichlet boundary condition for incoming neutrinos from at each boundary position, while the free boundary one is adopted for escaping neutrinos from the computational region. We run each simulation up to $10^{-4} {\rm ms}$.

\subsection{Classical simulations with BGK subgrid model}\label{subsec:numsetBGKSM}

The corresponding equation with our BGK subgrid model to Eq.~\ref{eq:BasEq_SpheQKE} can be written as,
\begin{equation}
  \begin{split}
& \frac{\partial \barparena{f}}{\partial t}
+ \frac{1}{r^2} \frac{\partial}{\partial r} ( r^2 \cos \theta_{\nu}  \barparena{f} )  - \frac{1}{r \sin \theta_{\nu}\
} \frac{\partial}{\partial \theta_{\nu}} ( \sin^2 \theta_{\nu} \barparena{f}) \\
&  = - \frac{1}{\tau_a} ( \barparena{f} - \barparena{f^a}),
  \end{split}
\label{eq:BGK_sphe}
\end{equation}
while we assume that the off-diagonal terms are zero, implying that Eq.~\ref{eq:BGK_sphe} is equivalent to classical neutrino transport. For this simulation, we extend our GRQKNT code \cite{2022PhRvD.106f3011N} by adding the BGK subgrid module. This exhibits that these numerical simulations for both full quantum kinetics and this classical Boltzmann transport with subgrid model have the same accuracy of neutrino advection. The initial- and boundary conditions are also the same as those in QKE simulations. In this demonstration, we employ Eq.~\ref{eq:approxiGrowth_NaMo} to estimate $\tau_a$. We note that $\tau_a$ is updated at every time step during the simulation.

To determine $\barparena{f^a}$, we employ a method in \cite{2023PhRvD.107j3022Z}. This offers an approximate scheme to determine asymptotic states of FFC analytically. As we shall discuss later, however, this analytic method corresponds to the simplest prescription and there is room for improvements. In fact, the scheme is developed based on assumptions that the neutrinos flight directions are $v>0$ (or $v<0$) and there is a single ELN angular crossing. These assumptions are not appropriate in general, leading to a systematic error in realistic situations. Nevertheless, this scheme can capture the essential trends of FFCs (see below), which may provide sufficient accuracy as a subgrid model.

In this method, we first compute the positive and negative ELN-XLN number densities,
\begin{equation}
\begin{split}
A \equiv & \left| \int_{G_{v}<0} d\Gamma G_{v}   \right|, \\
B \equiv & \int_{G_{v}>0} d\Gamma G_{v},
\label{eq:ABdef}
\end{split}
\end{equation}
while $G_v$ is given in Eq.~\ref{eq:Gamma_and_Gv}.
In cases with $B>A$ (positive ELN-XLN density), we determine $\eta$ in Eq.~\ref{eq:trapro_three} as,
\begin{equation}
\eta =
\begin{cases}
\frac{1}{3}  & ( G_{v}<0 )\\
1 - \frac{2A}{3B}  & ( G_{v} \ge 0 )
\end{cases}
,
\label{eq:survProb_BgtA}
\end{equation}
meanwhile $\eta$ in $B<A$ (negative ELN-XLN density) is determined as,
\begin{equation}
\eta =
\begin{cases}
\frac{1}{3}  & ( G_{v}>0 )\\
1 - \frac{2B}{3A}  & ( G_{v} \le 0 )
\end{cases}
.
\label{eq:survProb_BltA}
\end{equation}
In the case with $B=A$, $\eta$ is set to be $1/3$ for all $v$, indicating that $f^a$ corresponds to the complete flavor equipartition (see also \cite{2021PhRvD.104j3003W,2023PhRvD.107j3022Z,2023PhRvD.108f3003X}). We also note that $\bar{\eta}$ is equal to $\eta$, since we do not have to distinguish neutrinos and antineutrinos in FFC (see also \cite{2023PhRvD.107j3022Z}).

\begin{figure*}
   \includegraphics[width=\linewidth]{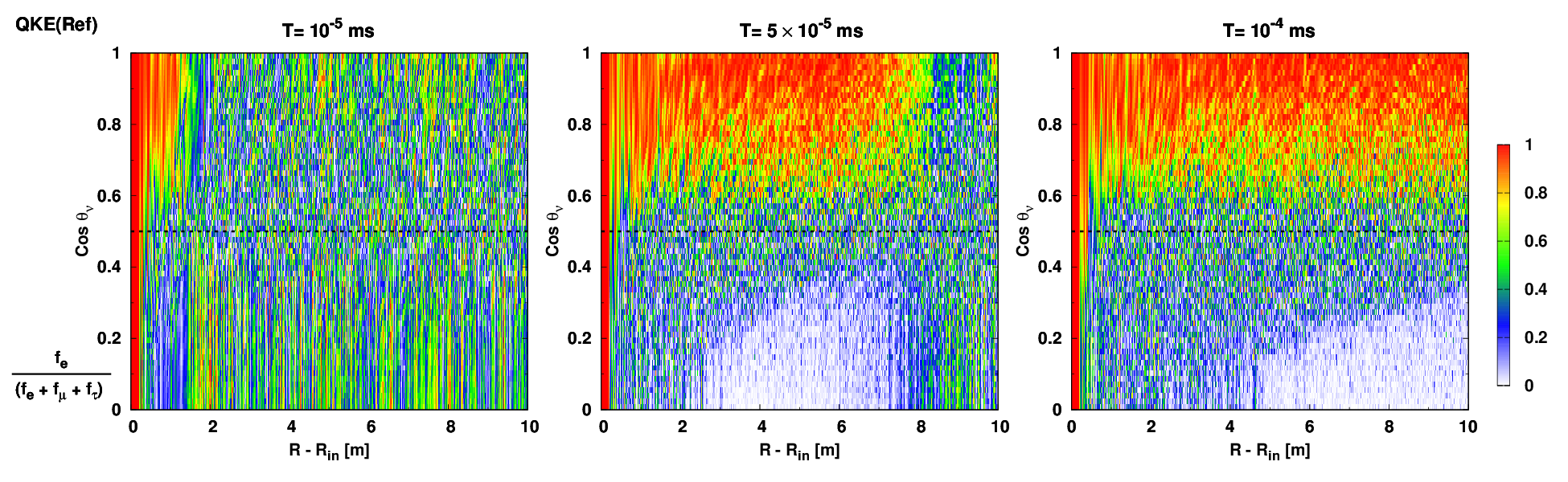}
   \includegraphics[width=\linewidth]{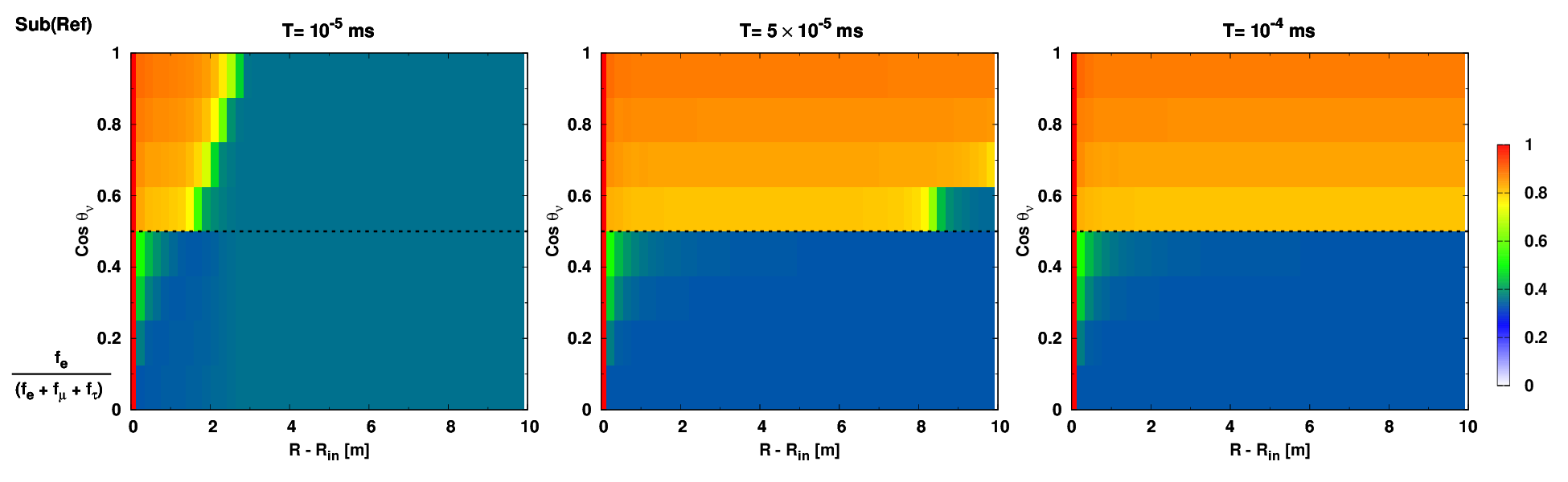}
   \caption{Color map of survival probability of $\nu_e$ for reference model ($\alpha=1$ and $\bar{\beta}_{ee}=1$). The horizontal and vertical axes denote radius ($R - R_{\rm in}$) and directional cosine of neutrino flight angle ($\cos \theta_{\nu}$), respectively. The dashed line in each panel represents the neutrino angle with ELN zero crossing at the initial condition. Top and bottom panels distinguish results by quantum kinetic simulation and classical one with BGK subgrid model. From left to right, we show the results at three different time snapshots: $T = 10^{-5}, 5 \times 10^{-5}$, and $10^{-4}$ms.
}
   \label{fig:graph_QKEsubcomp_A1B1}
\end{figure*}

\begin{figure*}
   \includegraphics[width=\linewidth]{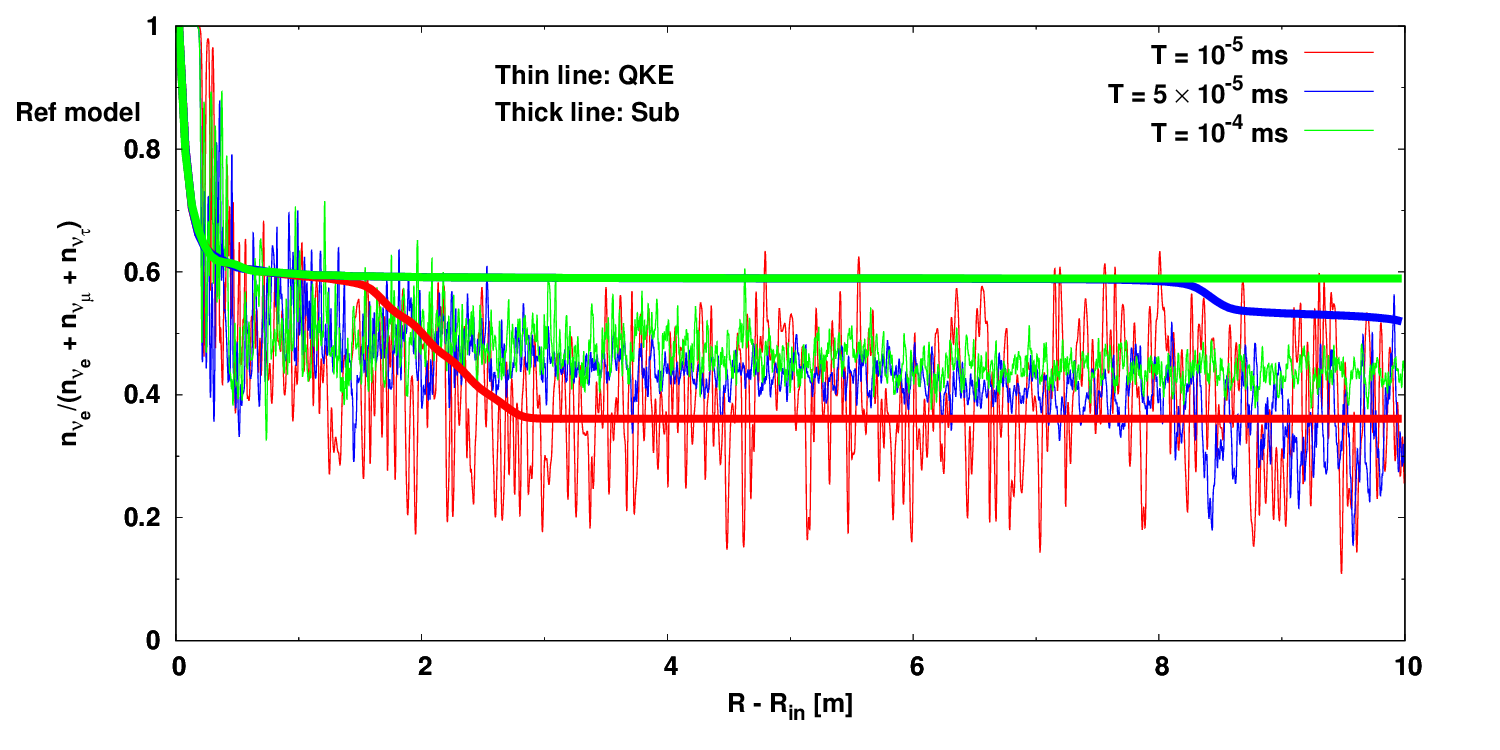}
   \caption{The radial profile of angular-averaged survival probability of $\nu_e$ for the reference model. The color distinguishes time snapshot. The thin and thick lines show results for quantum kinetic simulation and classical one with BGK subgrid model, respectively.
}
   \label{fig:graph_Rad_NumdenDeg_Ref}
\end{figure*}

\begin{figure*}
   \includegraphics[width=\linewidth]{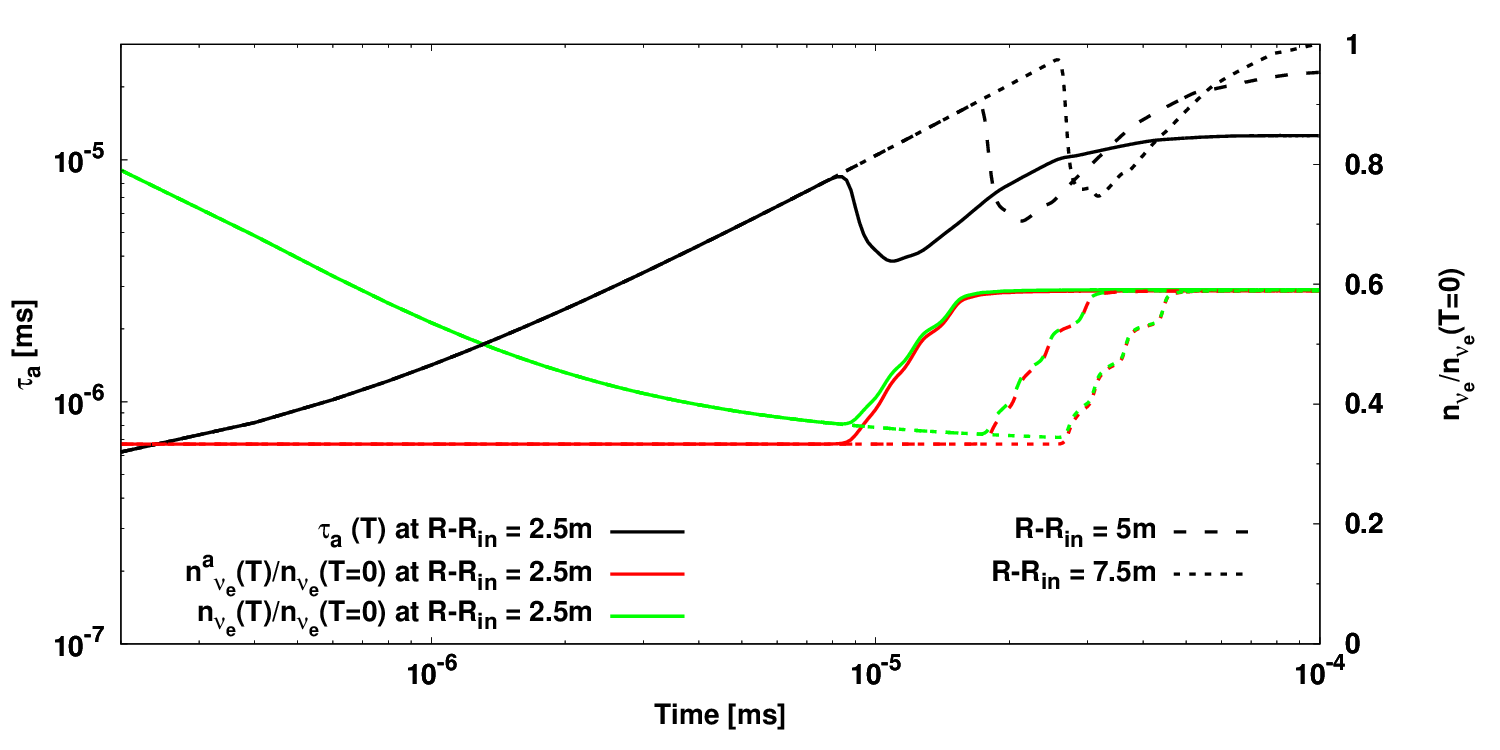}
   \caption{Time evolution of $\tau_a$ (black), $n^a_{\nu_e}$ (red), and $n_{\nu_e}$ (green) at three spatial positions for reference model. Line type distinguishes spatial positions: $R-R_{\rm in}=2.5$m (solid), $5$m (dashed), and $7.5$m (dotted). Left and right y-axes are for $\tau_a$ and $n^a_{\nu_e}$ ($n_{\nu_e}$), respectively. In this plot, $n^{a}_{\nu_e}$ and $n_{\nu_e}$ are normalized by $n_{\nu_e}$ at $T=0$ms.
See the text for more details.
}
   \label{fig:graph_tevo_fa_taua_A1B1}
\end{figure*}

Let us put an important remark here. As shown in \cite{2023PhRvD.107l3021Z}, the asymptotic state of FFCs obtained from quantum kinetic simulations depends on boundary conditions. In fact, the Dirichlet boundary condition (as used in this demonstration) results in qualitatively different asymptotic state from those obtained by periodic one. In the Dirichlet case, the asymptotic state is determined so as to preserve ELN- and XLN- number fluxes. In this demonstration, however, we determine $\eta$ from the condition of number conservation (Eqs.~\ref{eq:ABdef}-\ref{eq:survProb_BltA}), despite employing the Dirichlet boundary condition. One may wonder if this is inconsistent treatment. As we shall demonstrate below, however, our choice is appropriate. We will provide this detailed discussion in Sec.~\ref{subsec:results}.

One of the advantages of subgrid model is that high resolutions are no longer necessary in these simulations, since there are no driving terms to create small scale structures in this coarse-grained model. For this reason, we employ $N_r = 192$ and $N_{{\theta}_{\nu}}=16$ grid points with the same domains as those used in QKE simulations. It should be mentioned, on the other hand, that $\tau_a$ is much smaller than the advection timescale (which is also associated with Courant-Friedrich-Levy condition for the stability of numerical simulations), implying that Eq.~\ref{eq:BGK_sphe} becomes a stiff equation. This requires an implicit time evolution to numerically stabilize in solving the equation. In this demonstration, an operator-splitting approach is adopted, in which we first evolve $f$ by neutrino advection in time explicitly, and then the BGK term (right hand side of Eq.~\ref{eq:BGK_sphe}) is handled by an implicit way. More specifically, the distribution function of neutrinos at $n+1$ time step ($f^{n+1}$) is computed as,
\begin{equation}
f^{n+1} = (\frac{1}{\Delta t} + \frac{1}{\tau_a})^{-1} \biggl(
\frac{f^{*}}{\Delta t} + \frac{f^a}{\tau_a}
\biggr),
\label{eq:operatorsplit}
\end{equation}
where $\Delta t$ denotes the time step. In this expression, $f^{*}$ corresponds to a tentative distribution function which is obtained by $f$ evolved only by advection terms in Eq.~\ref{eq:BGK_sphe}. We confirm that this operator-splitting method works well to evolve the system in a numerically stable manner.

For the sake of completeness, a resolution study is also undertaken with reference model ($\alpha=1$ and $\bar{\beta}_{ee}=1$) of subgrid model. One of them is a simulation with twice higher spatial resolution than the reference one (i.e., $N_r = 384$), while the angular resolution remains the same. We also carry out another simulation with high angular resolution, $N_{{\theta}_{\nu}}=128$, which corresponds to the same resolution as that adopted in quantum kinetic transport, while the spatial resolution is the same as reference one ($N_r = 192$). As shown below, these results are essentially the same as reference model, exhibiting that simulations employed BGK subgrid models are not sensitive to numerical resolutions.

\subsection{Results}\label{subsec:results}

\begin{figure*}
   \includegraphics[width=\linewidth]{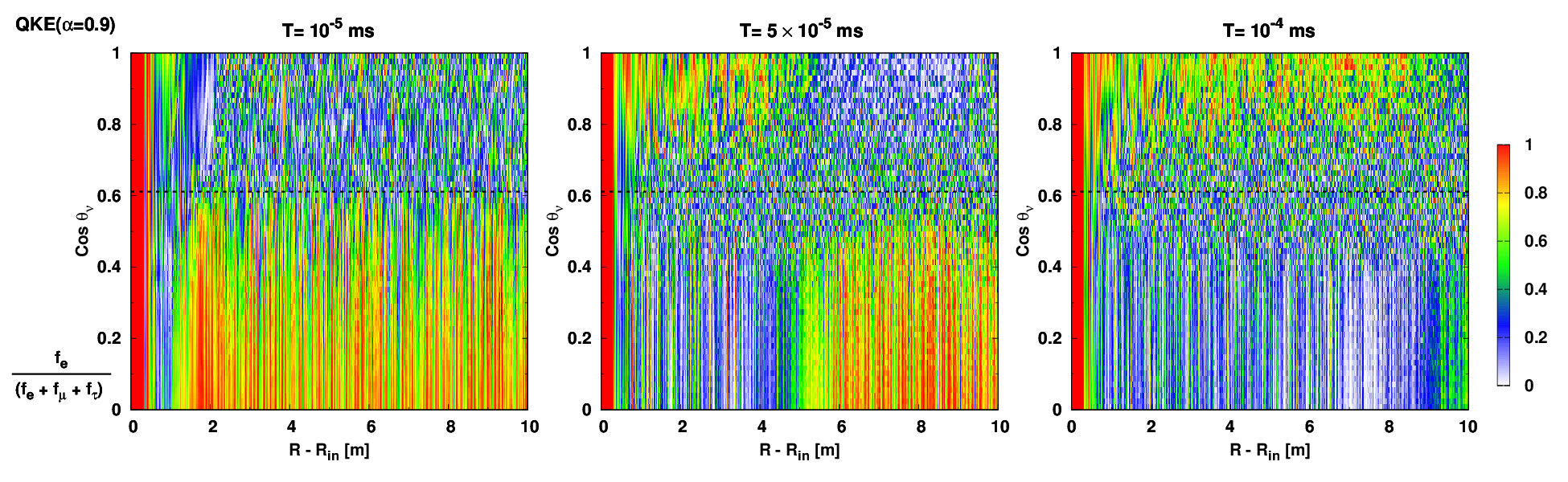}
   \includegraphics[width=\linewidth]{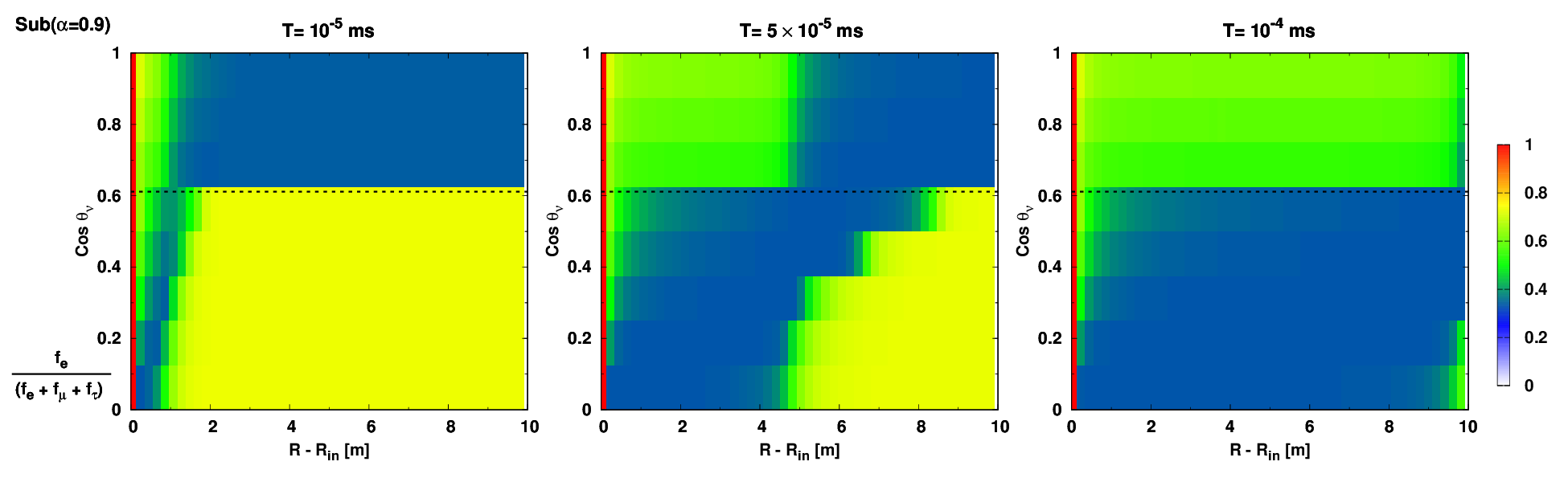}
   \caption{Same as Fig.~\ref{fig:graph_QKEsubcomp_A1B1} but for the model with $\alpha=0.9$ (and $\bar{\beta}_{ee}=1$).
}
   \label{fig:graph_QKEsubcomp_A09B1}
\end{figure*}

\begin{figure*}
   \includegraphics[width=\linewidth]{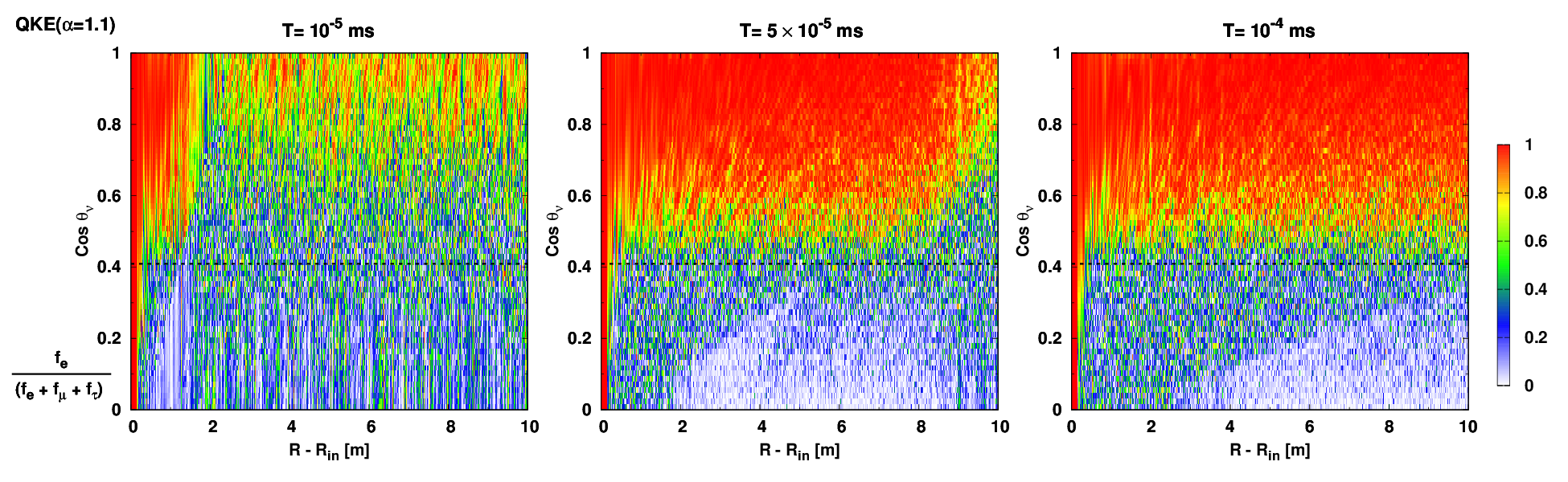}
   \includegraphics[width=\linewidth]{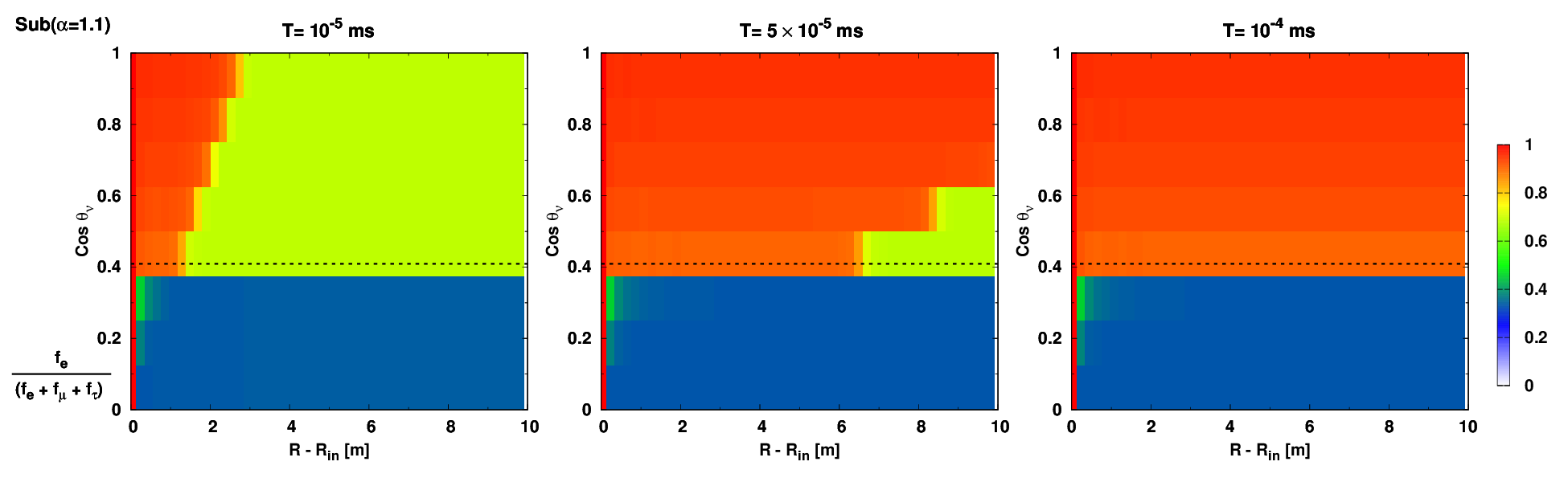}
   \caption{Same as Fig.~\ref{fig:graph_QKEsubcomp_A1B1} but for the model with $\alpha=1.1$ (and $\bar{\beta}_{ee}=1$).
}
   \label{fig:graph_QKEsubcomp_A11B1}
\end{figure*}

\begin{figure*}
   \includegraphics[width=\linewidth]{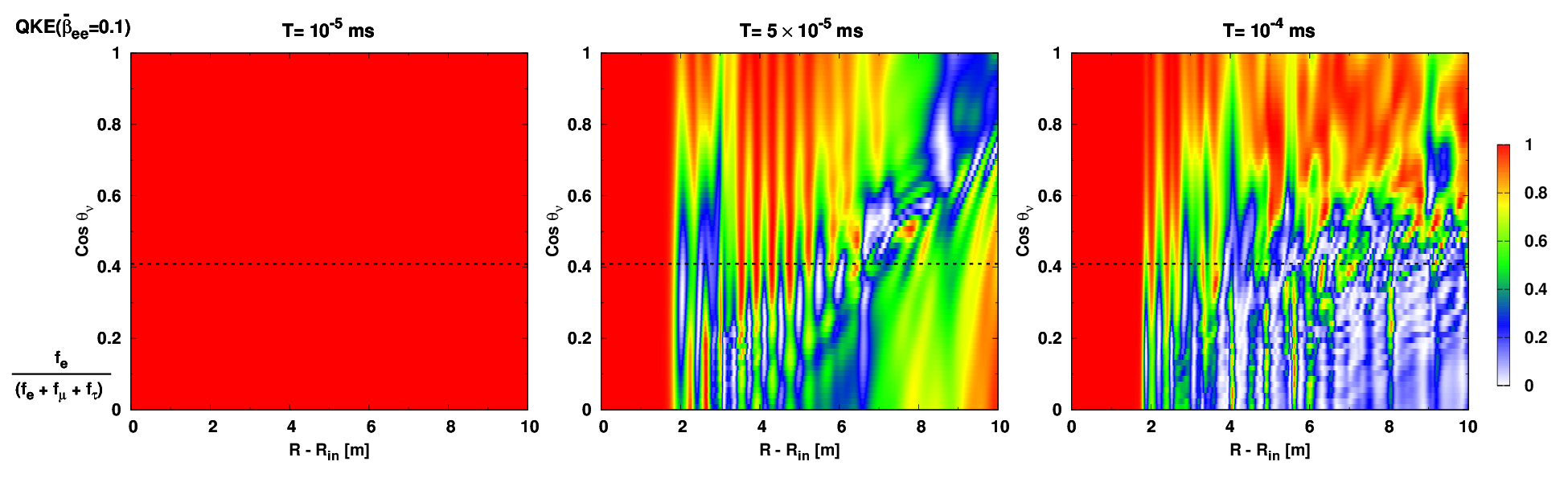}
   \includegraphics[width=\linewidth]{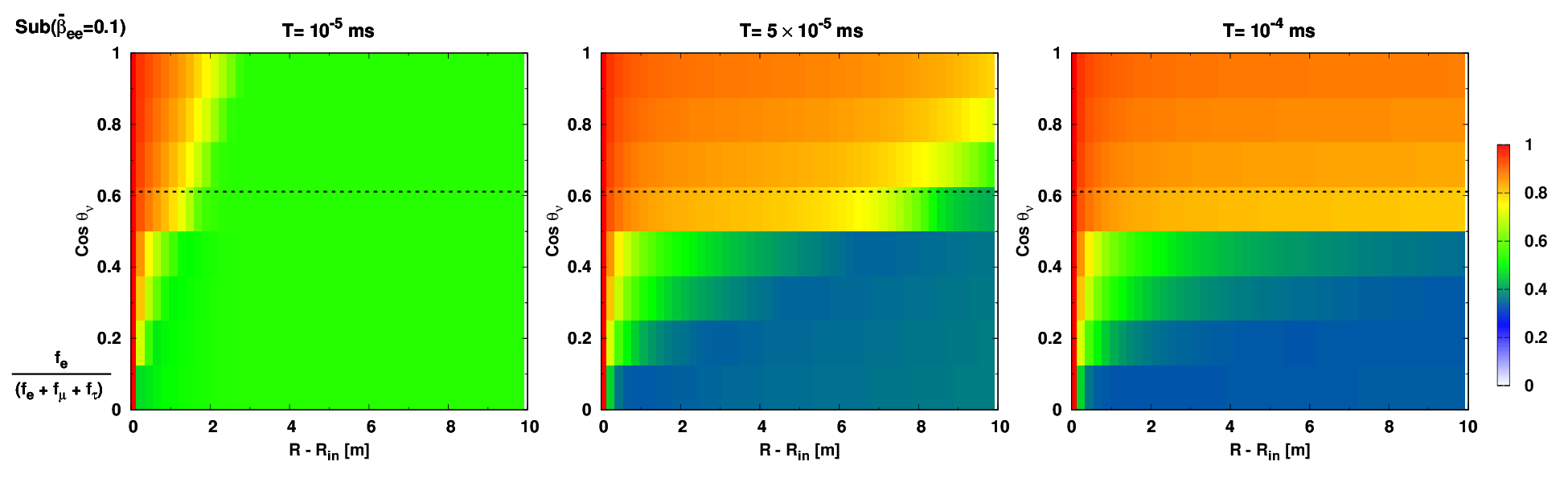}
   \caption{Same as Fig.~\ref{fig:graph_QKEsubcomp_A1B1} but for the model with $\bar{\beta}_{ee}=0.1$ (and $\alpha=1.0$).
}
   \label{fig:graph_QKEsubcomp_A1B01}
\end{figure*}

\begin{figure*}
   \includegraphics[width=\linewidth]{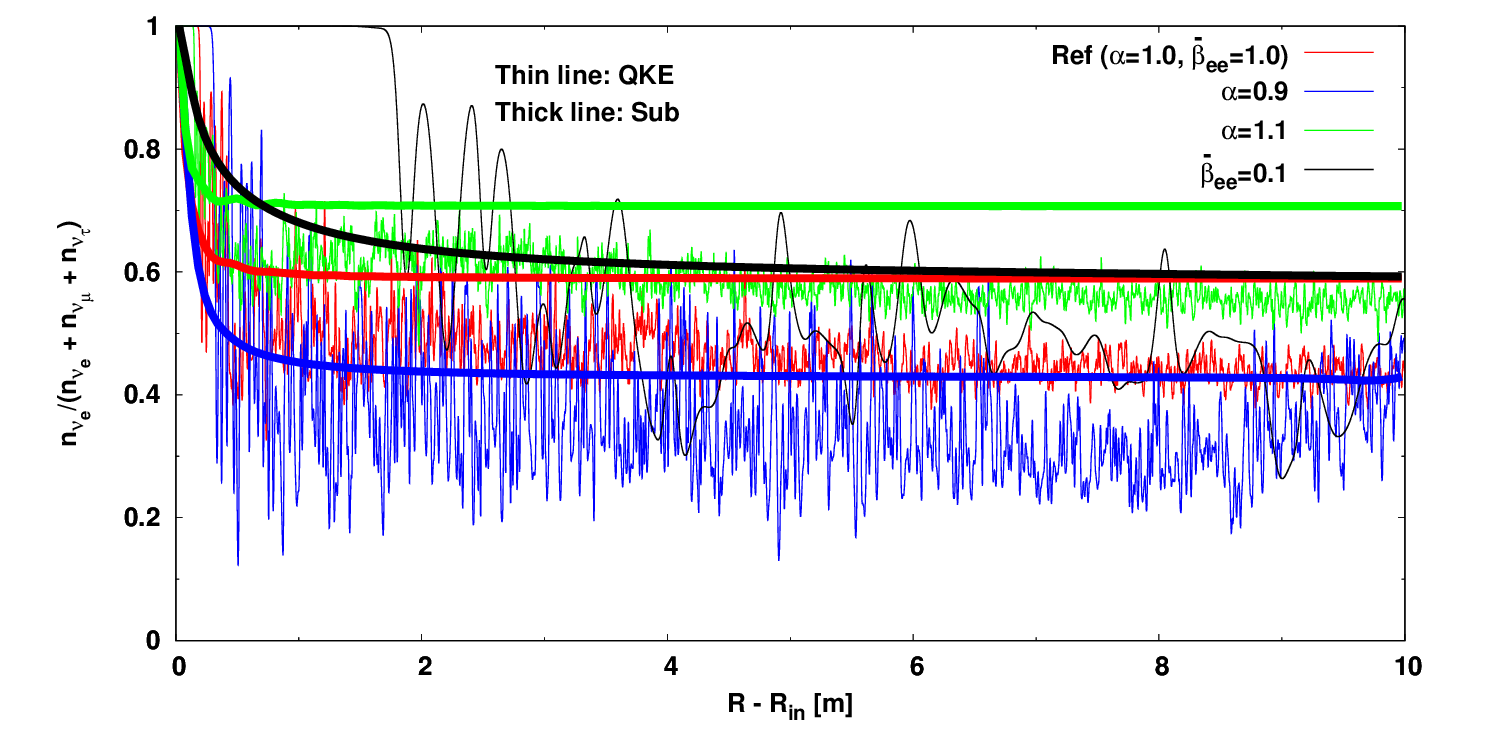}
   \caption{Same as Fig.~\ref{fig:graph_Rad_NumdenDeg_Ref} but for all models. The color distinguishes models. We display the results only at the end of simulation ($T = 10^{-4}$ms).
}
   \label{fig:graph_Rad_NumdenDeg_finalsnap}
\end{figure*}


\begin{figure*}
   \includegraphics[width=\linewidth]{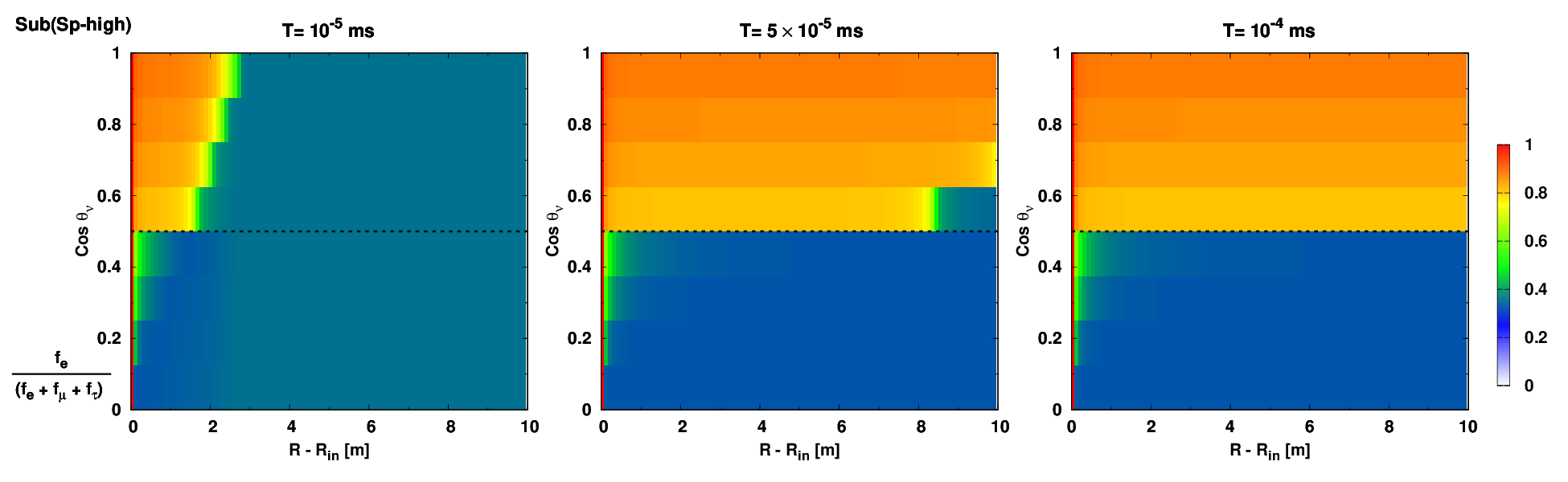}
   \caption{Same as Fig.~\ref{fig:graph_QKEsubcomp_A1B1} but for a subgrid model (reference model) with twice higher spatial resolution ($N_{r}=384$) than reference one.
}
   \label{fig:graph_sub_A1B1_sphigh}
\end{figure*}

\begin{figure*}
   \includegraphics[width=\linewidth]{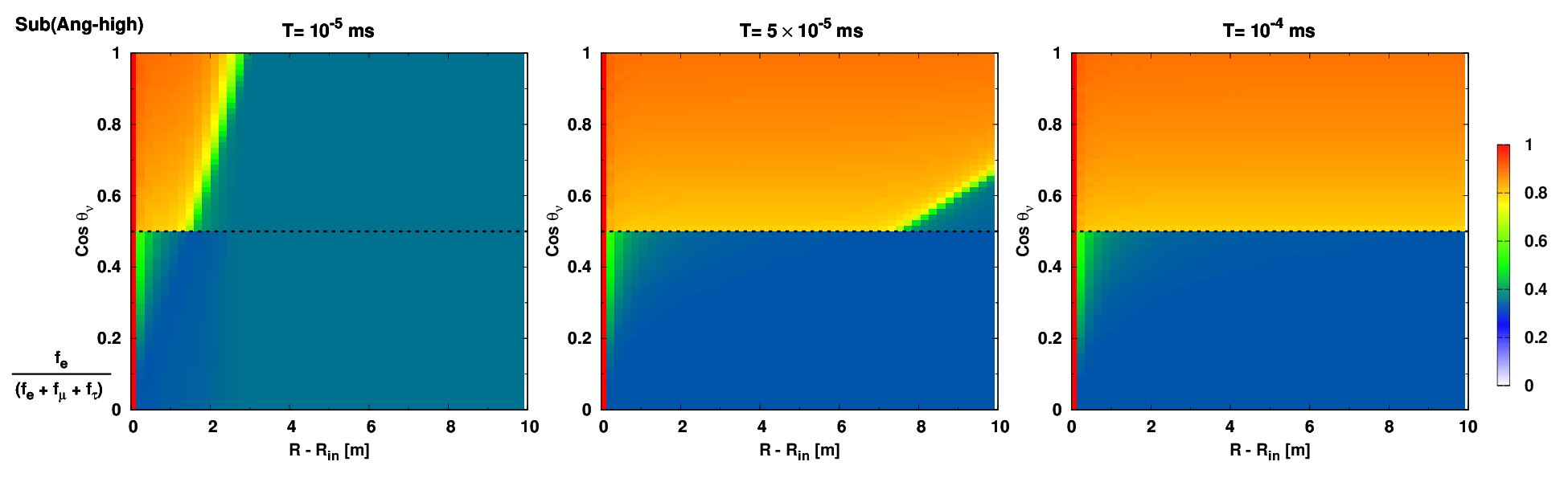}
   \caption{Same as Fig.~\ref{fig:graph_QKEsubcomp_A1B1} but for a subgrid model (reference model) with higher angular resolution ($N_{\theta_{\nu}}=128$) than reference one.
}
   \label{fig:graph_sub_A1B1_anghigh}
\end{figure*}


In Fig.~\ref{fig:graph_QKEsubcomp_A1B1}, we show the color map of survival probability of $\nu_e$ as functions of $r$ and $\cos {\theta_{\nu}}$. From left to right, results with three different time snapshots are displayed ($T = 10^{-5}, 5 \times 10^{-5}$, and $10^{-4}$ms, respectively). Top and bottom panels distinguish quantum kinetic model and classical one with BGK subgrid model. Since antineutrinos have essentially the same properties as those in neutrinos, we omit to show them.

As shown in the top left panel of Fig.~\ref{fig:graph_QKEsubcomp_A1B1}, neutrino flavor conversions vividly occur and reach nearly flavor equipartition in the almost entire neutrino flight directions at $T=10^{-5}$ms. This is consistent with the previous studies \cite{2021PhRvD.104j3003W,2023PhRvD.107j3022Z,2023PhRvD.108f3003X} that FFC makes the system evolve toward the flavor equipartition in the case with $n_{\nu_e} = n_{\bar{\nu}_e}$. As we discussed in \cite{2023PhRvD.107f3033N,2023PhRvD.107l3021Z}, however, the flavor equipartition is not the actual asymptotic state in cases with Dirichlet boundary condition. In fact, angular distributions of survival probability of $\nu_e$ become remarkably different around the boundary at $R-R_{\rm in}=0$; indeed, FFC tends to be less vigorous in $\cos \theta_{\nu} \gtrsim 0.5$. The region expands with time, and eventually it dominates the entire computational domain (see the top middle and right panels in Fig.~\ref{fig:graph_QKEsubcomp_A1B1}). We will discuss the physical mechanism of the transition in detail later, which is associated with the determination of $f^a$ from $f$ in BGK subgrid model.

As shown in the bottom panels of Fig.~\ref{fig:graph_QKEsubcomp_A1B1}, the corresponding classical simulation with BGK subgrid model can reproduce qualitatively similar results as those found in the quantum kinetic simulation. In the earlier phase, FFC occurs in the entire angular regions except for the vicinity of $R-R_{\rm in}=0$, but the flavor conversion in $\cos \theta_{\nu} \gtrsim 0.5$ subsides after neutrinos injected (constant in time) at $R-R_{\rm in}=0$ reach there. In Fig.~\ref{fig:graph_Rad_NumdenDeg_Ref}, we compare the radial profiles of the angular-averaged survival probability of $\nu_e$ between the two simulations, and we confirm that the errors are within $\sim 20 \%$. This comparison lends confidence to the capability of BGK subgrid model.

It is interesting to inspect how $\tau^a$ and $f^{a}$ vary in space and evolve with time. To see their essential features, we show the time evolution of $\tau^a$ and $n^{a}_{\nu_e}$ (the number density of electron-type neutrinos computed from $f^{a}_{ee}$) at three different radii ($R-R_{\rm in}=2.5, 5,$ and $7.5$m) in Fig.~\ref{fig:graph_tevo_fa_taua_A1B1}. As a reference, we also show $n_{\nu_e}$ in the same figure. In the early phase ($T \lesssim 10^{-5}$ms), $\tau^a$ monotonically increases with time, while $n_{\nu_e}$ approaches $n^{a}_{\nu_e}$. These time evolutions are identical among three different radii, indicating that the system evolves nearly homogeneously. The increase of $\tau_a$ exhibits that ELN-XLN angular crossings become shallow (see also Eq.~\ref{eq:approxiGrowth_NaMo}) due to $f \rightarrow f^a$. At $T \sim 10^{-5}$ms, the time evolution of both $\tau^a$ and $f^{a}$ becomes qualitatively different from that in the earlier phase. This phase corresponds to the transition of asymptotic states from periodic case to Dirichlet one. In fact, the onset timing of the phase transition is earlier for smaller radius, which exhibits that impacts of Dirichlet boundary condition propagate in the positive radial direction. During the transition phase, both $\tau^a$ and $n^{a}_{\nu_e}$ are dynamically evolved and also inhomogeneous in space, meanwhile $n_{\nu_e}$ keep approaching $n^{a}_{\nu_e}$. At $T \sim 10^{-4}$ms, the system settles into a steady state. Interestingly, $\tau_a$ remains finite at different positions and it varies with radius even at the end of our simulation. This exhibits that ELN-XLN angular crossings do not disappear completely in the steady state, and that the depth of ELN-XLN angular crossing is deeper for smaller radii. This trend can be interpreted by effects of neutrino advection under Dirichlet boundary condition. The neutrinos having ELN-XLN angular crossings are injected constantly in time at $R=R_{\rm in}$, and the angular crossing fades with radius. Nevertheless, $\tau_a$ is much larger than the time scale of neutrino-self interactions at $R \gg R_{\rm in}$, indicating that ELN-XLN angular crossings almost disappear.




Although the overall properties can be well captured by the BGK subgrid model, there are quantitative deviations, the origins of which are worth to be discussed. In the early phase, the growth of flavor conversion is slightly faster for the classical simulation with BGK subgrid model. This error comes from the empirical determination of $\tau_{a}$ by Eq.~\ref{eq:approxiGrowth_NaMo}, which does not have the ability to determine the growth rate of FFC quantitatively. We also find that some detailed angular-dependent features are not captured by the subgrid model. In quantum kinetic simulations, flavor conversions vividly occur in the region of $0 \le \cos \theta_{\nu} \lesssim 0.6$, but the angular region is slightly narrower for the subgrid model ($0 \le \cos \theta_{\nu} \lesssim 0.5$). This is mainly due to the accuracy of determination of $\eta$ in our subgrid model. As described in Eqs.~\ref{eq:survProb_BgtA}~and~\ref{eq:survProb_BltA}, the angular distribution of $\eta$ is discrete at $G_{v}=0$ in our approximate scheme, but it is continuous in real. Regarding this issue, one can reduce the error if we employ smooth functions to determine angular distributions of $\eta$, although the numerical cost may become more expensive. We note that such approximate schemes have been recently proposed by \cite{2023PhRvD.108f3003X}, and they showed that the quadratic functions can reduce the error by $30$ to $50 \%$ from our box-like treatment.

In Figs.~\ref{fig:graph_QKEsubcomp_A09B1}-\ref{fig:graph_QKEsubcomp_A1B01},
 we show the same plots as in Fig.~\ref{fig:graph_QKEsubcomp_A1B1} but for different models. These figures exhibit that the BGK subgrid model works well for all cases. One may think that the error around the boundary of $R-R_{\rm in}=0$ in the model with $\bar{\beta}_{ee}=0.1$ is higher than other models. However, this error is also due to the low accuracy of determining $\tau_a$, it can be improved if we employ better methods to determine it, for instance, based on linear stability analysis. In Fig.~\ref{fig:graph_Rad_NumdenDeg_finalsnap}, we compare the angular-averaged survival probabilities of $\nu_e$ at the end of our simulations among different models. For all models, we confirm that the error is within $\sim 20 \%$ for the asymptotic distribution of neutrinos.

We show the result of our resolution study in Figs.~\ref{fig:graph_sub_A1B1_sphigh}~and~\ref{fig:graph_sub_A1B1_anghigh}, that corresponds to the same plot as in Fig.~\ref{fig:graph_QKEsubcomp_A1B1}. As can be seen in these figures, the overall features are essentially the same as reference model, that lends confidence that the BGK subgrid model is applicable to numerical simulations with coarse resolutions.

Finally, we describe the reason why our BGK subgrid model with a prescription of Eqs.~\ref{eq:ABdef}-\ref{eq:survProb_BltA} works well, despite the fact that the flavor conversions in cases with Dirichlet boundary are qualitatively different from the periodic one. We start with discussing the mechanism of transition of asymptotic states from periodic case to Dirichlet one. As shown above, we observed at least temporarily in the early non-linear FFC phases that the asymptotic states determined based on the number conservation (i.e., periodic boundary case) appear in almost entire spatial region. This is because the dynamics of flavor conversions is almost identical in adjacent spatial regions, which offers the similar environment as a periodic boundary condition. As a result, the neutrino flux is also constant in adjacent spatial positions, guaranteeing the ELN- and XLN number conservation at each spatial position. On the other hand, both ELN and XLN number fluxes (or first angular moments) in this (temporal) asymptotic state become different from those in initial conditions, whereas they are fixed in time at the boundary of $R=R_{\rm in}$ due to Dirichlet condition. This is a crucial problem for asymptotic states, since the number flux needs to be balanced to achieve the steady state (see Eq. 6 in \cite{2023PhRvD.107l3021Z}). This implies that the neutrino distributions in the periodic boundary condition does not satisfy the actual asymptotic state. This also exhibits that ELN- and XLN number fluxes at $R>R_{\rm in}$ is different from $R=R_{in}$, resulting in evolving ELN- and XLN- number densities (or zeroth angular moments) at each spatial position.

One thing we do notice along this discussion is that the classical simulation with BGK subgrid model has the capability to handle the effects of neutrino advection precisely, since the advection term is the same as that in quantum kinetic one. This indicates that the dynamical evolution of ELN and XLN number densities at all spatial positions are well modeled. This also suggests that the neutrino radiation field obtained in the subgrid model evolves in time so that neutrino fluxes become constant in space to achieve the steady state, while this results in the dynamical change of ELN and XLN number densities. In BGK subgrid model, we determine $f^a$ by the time- and spatial dependent $f$ to satisfy ELN and XLN number density at each position, which leads eventually to the consistent asymptotic state determined from the conservation of ELN- and XLN number flux. This corresponds to the asymptotic state with Dirichlet boundary condition.

The above argument exhibits that the local study of flavor conversions with periodic boundary conditions is worthy to improve the BGK subgrid model. As demonstrated in \cite{2023PhRvL.130u1401N,2023PhRvD.108l3003N,2023PhRvD.108j3014N,2023PhRvD.107f3025S}, global advection of neutrinos affects the dynamics of flavor conversion significantly, and that the final outcome of neutrino radiation fields are qualitatively different from those estimated from local simulations. The present study suggests, however, that the effects of global advection can be decoupled from local dynamics of flavor conversion under the framework of our BGK subgrid model. This suggests that the classical BGK model has the capability of modeling global quantum kinetics of neutrinos in CCSN and BNSM environments by precise determination of $f^a$ and $\tau_a$ based on local study of flavor conversions.

\section{Summary}\label{sec:summary}
In this paper, we present a new subgrid model for neutrino quantum kinetics, in particular  for neutrino flavor conversion. The basic assumption in this subgrid model is to handle the dynamics of flavor conversions as a relaxation process, in which the flavor conversion makes the system to asymptotic states ($f^a$) in the time scale of $\tau_a$. This treatment is essentially the same as a BGK relaxation-time approximation \cite{1954PhRv...94..511B}, which was originally developed to approximately handle collisional processes in gas dynamics. In our model, we do not apply the approximation to collision term but neutrino oscillation term. We describe the QKE with the BGK model in Sec.~\ref{sec:basiceq}, and also provide an explicit form for two-moment method in Sec.~\ref{sec:twomoment}. We also present a concrete example of how we can use the BGK model in classical neutrino transport by focusing on FFC (Sec.~\ref{sec:demo}). We assess the capability of the BGK subgrid model by comparing to the results of quantum kinetic neutrino transport, and show that the subgrid model has the ability to capture the overall features in dynamics of neutrino flavor conversions.

Although our subgrid model is a valuable tool with many potentials, more work is certainly needed to increase the accuracy. It should be pointed out that the present study also provides a strategy to improve the subgrid model. As shown in Eq.~\ref{eq:trapro_three}, accurate determination of $\eta$ (and $\bar{\eta}$) from $f$ is crucial and any approaches including analytic schemes \cite{2023PhRvD.107f3033N,2023PhRvD.107j3022Z,2023PhRvD.107l3021Z,2023PhRvD.108f3003X} and AI \cite{2023arXiv231115656A} are applicable. We note that the prescription that used in the present demonstration (see Eqs.~\ref{eq:ABdef}~to~\ref{eq:survProb_BltA}) is just an example for FFC, but we certainly need others for different types of flavor conversions. In fact, the analytic scheme with Eqs.~\ref{eq:ABdef}~to~\ref{eq:survProb_BltA} can not handle a flavor swap phenomena recently found in FFC simulations of BNSM environments \cite{2023PhRvD.108j3014N,2023arXiv231113842Z}. As such, we still need to improve approximate schemes to determine asymptotic states of FFCs.

We are also interested in how well the BGK subgrid model can work in cases that flavor conversions and collision processes (neutrino emission, absorption, and scatterings) are interacted to each other. As demonstrated in \cite{2021PhRvD.103f3002S,2021PhRvD.103f3001M,2022PTEP.2022g3E01S,2022PhRvD.106l3013K,2022PhRvD.105d3005S,2022PhRvD.106j3031P,2023PhRvD.108b3006K}, the asymptotic states of flavor conversion depends on neutrino-matter interactions. The detailed study is necessary to assess the capability of our subgrid model in such complicated systems. The detailed study is postponed to future work.

Although there is certainly room for improvements, the BGK subgrid model is very useful and easy to be implemented into currently existing CCSN and BNSM codes. This indicates that the global neutrino-radiation-hydrodynamic simulations with  respectable physical fidelity of flavor conversions become feasible. We hope that the BGK subgrid model contributes to the entire CCSN and BNSM community to accommodate effects of neutrino quantum kinetics into their simulations.

\section{Acknowledgments}
We are grateful to David Radice for useful discussions. The numerical simulations are carried out by using "Fugaku" and the high-performance computing resources of "Flow" at Nagoya University ICTS through the HPCI System Research Project (Project ID: 220173, 220047, 220223, 230033, 230204, 230270), XC50 of CfCA at the National Astronomical Observatory of Japan (NAOJ), and Yukawa-21 at Yukawa Institute for Theoretical Physics of Kyoto University. For providing high performance computing resources, Computing Research Center, KEK, and JLDG on SINET of NII are acknowledged. This work is also supported by High Energy Accelerator Research Organization (KEK). HN is supported by Grant-inAid for Scientific Research (23K03468) and also by the NINS International Research Exchange Support Program. LJ is supported by a Feynman Fellowship through LANL LDRD project number 20230788PRD1. MZ is supported by a JSPS Grant-in-Aid for JSPS Fellows (No. 22KJ2906) from the Ministry of Education, Culture, Sports, Science, and Technology (MEXT) in Japan.
\bibliography{bibfile}

\end{document}